\documentstyle[12pt,preprint,epsf,epsfig,rotate]{aastex}

\newcommand{\be}{\begin{equation}}
\newcommand{\ee}{\end{equation}}

\begin{document}
\title{3D-MHD simulations of  an accretion disk
with star-disk boundary layer} 

\bigskip
\author{Adriane Steinacker}  

\affil{NASA Ames Research Center, Moffett Field, CA 94035}

\email{adriane@duras.arc.nasa.gov}

\author{John C.B. Papaloizou} 

\affil{Astronomy Unit, School of Mathematical Sciences, Queen Mary and Westfield College,
Mile End Road, London E1 4NS}

\email{J.C.B.Papaloizou@maths.qmw.ac.uk}

\begin{abstract}

We present global 3D MHD simulations of geometrically thin
but unstratified  accretion disks
in which a near Keplerian disk rotates between
two bounding regions with  initial rotation profiles that are stable
to the MRI. The inner region models the boundary layer between
the disk 
and an assumed more  slowly rotating central, non  magnetic star. 
We investigate the dynamical evolution of this system in response to
initial  vertical and toroidal fields imposed 
in a variety of domains contained within  the  near Keplerian disk. 
Cases with both  non zero and zero net magnetic  flux are considered
and  sustained dynamo activity
found in  runs for up to fifty orbital periods at the outer boundary
of the near Keplerian disk. 

We find a progression
of behavior regarding 
the turbulence resulting from the MRI and the 
evolving structure of
the disk and boundary layer
according to the initial field configuration. Simulations
starting  from  fields with small radial scale and
with zero net flux lead to the lowest
levels of turbulence and smoothest variation of disk mean state variables.
As  found in local simulations,
the final  outcome is shown to be independent of the form
of the imposed field.
 For our computational set up, average
values of the Shakura \& Sunyaev (1973) $\alpha$ parameter
in the Keplerian disk are  typically
$0.004\pm 0.002.$ 
Magnetic field eventually always diffuses into the boundary
layer  resulting in the build up of toroidal field,
 inward angular momentum transport and the accretion of
disk material. The mean radial velocity,
while exhibiting large temporal fluctuations
is always subsonic.

Simulations starting  with net toroidal  flux
may yield an average $\alpha \sim 0.04.$ 
While being characterized by one order of magnitude larger
average $\alpha$, simulations starting from
vertical fields with large radial scale and net flux may 
lead to the formation of persistent  non-homogeneous, non-axisymmetric
 magnetically dominated regions of very low density. 
In these gaps, 
angular momentum transport occurs through
 magnetic torques acting between regions on either side of the gap.
Local turbulent transport occurs where the  magnetic field is not dominant.
These simulations are indicative of the behavior
of the disk when  threaded by magnetic flux originating from an external source.
However, the  influence  of such presumed  sources  in determining
the
boundary conditions  that should be applied to the disk
remains to be investigated.

\end{abstract}


\section{Introduction}\label{S0} 
\noindent
The study of the boundary layer, the region where the angular
velocity of  an accretion disk drops to match the rotation velocity of the
central object, is of great importance for the understanding of accreting
objects, since up to half of the total accretion energy  can be released
there (Lynden-Bell $\&$ Pringle 1974)  if the central object
is a slow rotator.
 Up to now detailed studies of this region (eg. Papaloizou \& Stanley 1986, hereafter PS, Kley 1989,
Popham \& Narayan 1992)  have been
based on the Navier-Stokes equations  
with a modified viscosity prescription involving an anomalous viscosity coefficient.
 This is presumed to contain the effects of any turbulence present
reducing the problem to  one of
laminar flow. 

\noindent
The decrease of angular velocity in the boundary
layer is associated with an increase in the thermal pressure,
which
replaces  centrifugal support for the accreting  matter.
The large pressure gradients in turn, may be associated with supersonic
radial infall velocities in this region, if a standard disk viscosity
prescription is  continued into the layer. However, it has been argued that if such
a supersonic flow occurred, the star would loose its causal connection
to the outer disk (Pringle 1977).
In relation to this issue, several studies
based on the Navier-Stokes equations  were performed but
with a modified viscosity prescription in the boundary layer corresponding
to its much lower pressure scale height (PS,
Popham  \& Narayan 1992). This modification alone  does not eliminate
supersonic flows under all conditions (e.g. for large values
of $\alpha\simeq 1$). Popham  \& Narayan (1992)  suggested 
reducing the viscosity coefficient to zero as the radial infall
 velocity approaches
the  sound speed. This causally limited viscosity  (Narayan 1992)
always leads to subsonic infall.
A different approach,  appreciating the short time
required for matter to pass through the layer, allows the viscous stress
components to relax towards their equilibrium values on a relaxation
time scale (Kley \& Papaloizou 1997) and so naturally
incorporates causality. Studies of  one-dimensional models led to  
the conclusion that the boundary layer must be characterized not only
by the value of $\alpha$ in the outer disk, but also by the nature
of the viscous relaxation process. Additional time dependent studies
demonstrated that for low $\alpha\simeq 0.01$,
the boundary layer adjusts to a steady state, while for large
$\alpha=0.1$, significant disturbances occurred
in the boundary layer and to the power output. Periodic oscillations  
were seen throughout the disk by PS, whenever $\alpha$ was large (close
to 1). For small 
$\alpha$, the oscillations were localized near the outer disk boundary.
These oscillations are caused by the viscous overstability 
found by Kato (1978) and Blumenthal et al. (1984) 
and it was suggested they may be important in explaining the
time-dependent behavior in accreting objects such as CVs and protostars.

\noindent
All of the above models were based on an ad-hoc
anomalous viscosity  prescription  and  did not consider
further  its origin.
The discovery
of the relevance of the magnetorotational instability (MRI)
(Balbus  \& Hawley 1991)
has opened up a new era in  accretion disk
astrophysics. The instability provides a robust and self-consistent
mechanism for the production of turbulence and angular momentum
transport
in these objects if they are adequately ionized, thereby
removing the need for  ad-hoc prescriptions. 
The development of different numerical codes has enabled a
detailed investigation of the nonlinear phase of the instability. First
numerical studies were performed in a local shearing box approximation, 
(Hawley, Gammie \& Balbus  1995, 
Brandenburg et al. 1995, 
Sano, Inutsuka \& Miyama 1998, Fleming, Stone \& Hawley 2000, 
Miller  \& Stone 2000).
These studies show that the turbulent
outcome of the MRI depends on the initial field configuration applied
to the disk. Thus, local simulations with initial
vertical fields with
zero net flux field indicate an average  $\alpha=0.001-0.006$,
while vertical fields with non zero
net flux result in larger values of $\alpha$ up to $0.3.$
The turbulent outcome of an initially  unstable 
toroidal field can  lead to intermediate
values of $\alpha$ up to $ 0.04$
depending on the net flux.

\noindent
Relatively recently  studies of  instabilities in
global  disk models have begun (Armitage 1998). Such studies
are needed to see how
an unstable disk   modifies its  underlying  structure  in response to
globally varying levels of turbulence and whether the   
longer time scale
evolution  is in any way like that of
standard 
$'\alpha'$ disk  models.  Recent  studies 
have  been made by Hawley (2000), who concentrated on the evolution
of thick tori, and Hawley \& Krolik (2001). The latter study focuses
on the evolution
of the inner region of a disk 
accreting onto a black hole 
modeled with a pseudo-Newtonian potential.

\noindent
In this paper we study the interaction of an accretion disk
with a boundary layer region located between it and the central star. 
This situation is the relevant one to consider for
non relativistic accretion onto a non magnetic
central star. The inner boundary layer
region together with an exterior stable  region 
also provide convenient, relatively inert 
regions in which to embed a near Keplerian
disk with an unstable rotation profile. 
Instead of prescribing the viscosity
in an ad-hoc fashion, as in previous studies of the boundary layer,
we self-consistently incorporate the turbulence arising from
the MRI as the source of viscosity and diffusion of
magnetic field. 

\noindent
We assume the disk to have a small ratio of  scale height to radius
( $H/R\simeq 0.1$ ). 
The gravity is assumed to be entirely
due to the central object and for simplicity and in common
with other global disk studies  
we neglect vertical stratification  by  adopting  a cylindrically symmetric potential
thus focusing the study on the radial structure of the disk.
We study  the dynamical evolution  of the disk
over a time span of up to 
one thousand rotation periods  measured at the inner disk  edge. We consider 
different initial magnetic field configurations (poloidal and toroidal)
imposed in the main body of the disk.
Cases with both  non zero and zero net magnetic flux are considered.

\noindent
Simulations
starting with small scale initial fields with zero net flux
exhibit the lowest
Shakura  \& Sunyaev (1973)
parameter $\alpha$ with  a   mean value averaged over the
Keplerian domain  of 
$\simeq 0.005$. In this case
the simulations  on average attain a final state characterized by
the same mean $\alpha$ and  magnetic energy  independently of (within computationally defined limits)
the initial field strength.   There may also be a tendency
for  the mean value of 
$\alpha$ to increase with the extent of the vertical domain 
and the numerical resolution.  
This  is in agreement with results of
shearing box simulations (Hawley, Gammie \& Balbus 1996, hereafter HGB96). 
On the other hand simulations with large net  magnetic flux
may evolve turbulence with a larger 
mean value of $\alpha$ of $\simeq 0.04$, when the
initial field is toroidal.  Models starting from  initial 
vertical  fields with large radial scale
are such that $\alpha$ attains maximum values $>1$ in regions associated with prominent
density depressions, while outside these gaps, $\alpha$ reaches values similar to 
the zero net flux models (as low as $0.005).$ The volume averaged $\alpha$ depends on
the initial plasma beta in such cases.

\noindent We find that  all simulations locally  exhibit strong variations of
the vertically and azimuthally averaged values of  $\alpha$
in time and with  radius.  
All of the models display 
oscillations of the vertically and
azimuthally averaged  radial Mach number.
 Even though the boundary layer
region is stable to the MRI, magnetic field always diffuses into it.
This is the case even when the initial field is non zero only well
away from the layer. Toroidal field build up 
enables mass
to accrete through it onto the central star through the operation
of magnetic torques.

\noindent
The paper is organized as follows: 
in \S \ref{S1} we present the basic model and 
computational set up.  In \S \ref{S2}
we describe the numerical procedure.
In \S  \ref{S3} we discuss azimuthal and vertical
averaging together with  the global transport
of angular momentum and energy dissipation in the disk.
\S \ref{S4} is devoted to the investigation of  cases with
initial  
fields with zero net magnetic flux.
In \S \ref{S5} we present  results 
when the initial magnetic field  has net flux.
Finally, we summarize our results in \S \ref{S6} .

\section{Initial model setup} \label{S1} 
\noindent
The simulations are performed within the framework of ideal MHD. 
The governing equations
are:
\begin{equation}
\frac{\partial \rho}{\partial t}+ \nabla \cdot {\rho\bf v}=0, \label{cont}
\end{equation}
\begin{equation}
\rho \left(\frac{\partial {\bf v}}{\partial t}
 + {\bf v}\cdot\nabla{\bf v}\right)=
-\nabla p -\rho \nabla\Phi +
\frac{1}{4\pi}(\nabla \times {\bf B}) \times {\bf B}, \label{mot}
\end{equation}
\begin{equation}
\frac{\partial {\bf B}}{\partial t}=\nabla \times ({\bf v} \times {\bf B}).
\label{induct}
\end{equation}
where all the quantities have their usual meanings.
\noindent
While in some of our exploratory simulations 
we adopted an adiabatic equation of state 
(with heating due to artificial viscosity
retained), the 
runs described
in this paper are either
based on  using a locally isothermal equation of state, i.e.
\begin{equation}
P(R)=\rho(R)\cdot c(R)^2,
\end{equation}
where $c(R)$ denotes the sound speed in the disk
specified as a fixed function of $R,$
or after using this initially, solving the energy equation assuming that
heating  due to artificial viscosity is removed
by an exactly compensating cooling. Either of those
procedures gave very similar results.
We found them to  lead to a more
consistent procedure
since the numerical scheme currently applied with zero resistivity
does not allow for magnetic energy dissipated  
on the grid scale to be recovered
as heat. Tests showed that when
an adiabatic condition was used,
such that kinetic energy dissipated
through an artificial viscosity was recovered as heat,
there was a noticeable change in the thermal energy
content in the Keplerian disk during the course of a simulation. 

\noindent
The  computations are carried out  in cylindrical coordinates
 $(z,R,\phi)$. We assume that the gravitational
potential is only dependent on the radial coordinate, $\Phi=-GM/R$, where $M$ is the mass
of the central object and $G$ is the gravitational constant. 

\noindent
Our computational domain is divided into three distinct regions: 1) An extended active Keplerian domain,
ranging from a  radius $R_1$ to $R_2$,
in which the growth of the MRI leads to turbulence (Region I);
2) An inner boundary layer ranging from $R_0$, the lower radial boundary 
of the computational domain, to $R_1$ (Region II),
and 3) an outer region ranging from $R_2$ to
the outer radial boundary,
which we have added to avoid numerical problems
with trying to apply a boundary condition 
directly at the outer edge of the
active domain
(Region III). The corresponding values of $R_1, R_2$ and $R_3$ are specified
in table 1.
The initial density and angular velocity profiles 
for these three regions are calculated
consistently with hydrostatic equilibrium in the radial direction.
We present these solutions
below. 

\noindent
{\bf Region I: $R_1\le R < R_2$}.
In contrast to previously published
cylindrical global accretion disk simulations
(Armitage 1998, Hawley 2000 and Hawley \& Krolik 2001) 
where constant density
profiles were used, we adopt a 
 density varying inversely  with the radial distance
\begin{equation}
\rho(R)=\rho_0\cdot R_0/R,
\end{equation}
and a radial dependence
of the sound speed given by
\begin{equation}
c(R)=c_0 \sqrt{R_0/R} \label{sound}
\end{equation}
where $\rho_0$ and $c_0$ are the density and sound speed corresponding to the
radial position $R_0$. With
the pressure determined by a locally isothermal equation of state, we calculate
the rotation velocity
\begin{equation}
v_{\phi}(R)=\sqrt{G M/R_0-2c_0^2}\cdot \sqrt{R_0/R}\label{velc}.
\end{equation} 
For a thin disk in which the sound speed at a given radius is much smaller than the rotation
velocity, this profile is nearly Keplerian.

\noindent
{\bf Region II: $R_0\le R \le R_1$.}
In this region we require the angular velocity $\Omega$ to drop from its near-Keplerian value
at $R=R_1$ to a lower value $\Omega_\ast$ 
 which can be considered as matching  the angular velocity of a central stellar
object. For the rotation velocity, setting $i=1,$  we adopt
\begin{equation}
v_{\phi}(R)=v_{\phi i} \cdot (R/R_i)^{n_b},
\end{equation} 
and for the sound speed  we take
\begin{equation}
c(R)=c_i (R_i/R)^{n_c}.
\end{equation} 
These prescriptions lead to the following density profile:
\begin{equation}
\rho(R)=\rho_i\cdot(R/R_i)^{2n_c}\cdot\exp
{\left[\frac{0.5\delta -1}{n_b+n_c}\cdot\left((R/R_i)^{2(n_b+n_c)}-1\right)-\delta\cdot f_{i}(R)
\right]},
\end{equation}
where $f_{i}(R)=\ln(R/R_i)$, if $n_c=0.5$, and $f_{i}(R)=\left[(R/R_i)^{2n_c-1}-1\right]/(2n_c-1)$,
otherwise, and $\delta=GM/(R_0 c_0^2)$. Here $ \rho_i, c_i$ and $v_{\phi i}$ are the density,
sound speed, and rotation velocity,
at the radius $R_i$. These quantities are given by eq.\ (5), (6) and (7), respectively.
\noindent
We have considered cases with $n_b=1$ (corresponding to uniform
rotation), $n_b=2$
and $n_b=3.$ 
When $n_b=2,$ we  took  $n_c=0$.
 When $n_b=1,$ or $n_b=3$ we  took $n_c=0.625$ in most of the cases. Situations with $n_c=0.5$
have also been considered.
  We note that tests showed that  when this region was initiated with a density profile
not too far out of hydrostatic equilibrium,  the situation rapidly adjusted
so that   such an equilibrium was attained.

\noindent
{\bf Region III: $R_2 < R \le R_3$.}
This region is added primarily for numerical stability. It provides a high inertia
which prevents the magnetic field from diffusing into the outer radial boundary 
and producing a severe drop in the density and angular velocity of the 
boundary zones during the course of the simulations.
This  would ultimately  contaminate
the global energetics, leading to incorrect  behavior of the magnetic
energy and the $\alpha$ parameter.
\noindent
In this region, the solutions presented above for region II continue to be valid,
but with the index $i=2$.
In practice  we adopted  uniform initial rotation profiles
that are stable with respect to the MRI ($n_b=1$) and  $n_c=0.5.$

\noindent
The assumption of a thin disk requires the sound 
speed to be smaller than the rotation velocity,
which is equivalent to the pressure scale height $H=c/\Omega,$ the disk
would have if stratified vertically,
being significantly less than the current 
radius. We chose $c_0^2=0.01 GM/R_0$ for most models
but also ran two (b4,b5) with 
 $c_0^2=0.04 GM/R_0$ for comparison purposes.

\noindent
We comment that 
due to the chosen radial dependence of the sound speed and rotational velocity in region I, 
the scale height $H$ is a linear function of $R.$
In most cases  we chose the extent of the computational
domain in $z$ to be larger than one scale height at the outer
radial edge of the Keplerian domain. In few cases this was taken to be
one half of the scale height at the outer radial boundary of the Keplerian domain, thus
allowing us to test the effect of the extent of the $z$ domain on the results. 

\noindent
For most  calculations
with zero net flux  fields,
the radial domain was chosen such that $R_1=1.2, R_2 =3.7$ and $R_3=4.5.$
The $\phi$-domain extends from
0 to $\pi/2$ in most models. 
 Recent work by Hawley (2001) suggests that, with the same
resolution,  the extent of the $\phi$
domain does not  greatly affect the results. We have checked this by
performing runs with the $\phi$ domain
extending to 
$\pi$, and $\pi/3$. The boundary conditions in $z$ and $\phi$ are
periodic. In $R$ we set
the scalar quantities, magnetic field components and 
z- and $\phi$-components of the velocity
to have zero gradient.
The radial velocity components
at the  inner and outer  radial boundaries were set to zero. This 
ensures mass conservation in the computational domain.
 We also note that the computational set up is such that the 
zero gradient condition
on the azimuthal velocity applied at the radial boundaries  only affects values at ghost zones that do
not affect the flow elsewhere.
Thus artificially imposed  viscous boundary layer effects
do not occur in our calculations.

\noindent
We have performed simulations with:
\begin{enumerate}
\item
Initial vertical fields with and without net flux,
\item
Initial toroidal fields with   and without net flux,
\end{enumerate}
the magnetic field being initially defined in a restricted radial
domain:
\begin{equation}
{\bf B_i}=B_0\sin\left(2 n_R\pi \frac{R-R_b}{R_{b1}-R_b}\right)\bf{e_i}, \label{Binit}
\end{equation}
whith $n_R=0.5, 1.5$ for net field runs, and $n_R=3, 6, 9$ for runs with zero net flux.
The index $i$ indicates either the vertical or the toroidal field component
with the corresponding unit vector $\bf{e_i}.$  For toroidal fields
$B_0$ is a constant while for 
vertical fields $B_0 \propto 1/R,$
with $R_b$ and $R_{b1}$
being the boundaries of the region where the field was applied.
 These will be specified
for each run in turn.

\noindent
An overview of the performed simulations is given in table 1,
which shows the magnetic
field topology used, 
the  mean plasma beta defined as the ratio of thermal
to magnetic pressure with each of these
 averaged over the Keplerian
domain,
the numbers of computational grid points
in each  direction, the number of wavelengths of
the mode of maximum growth for the MRI contained in the z-domain
 ($z_u$ and $z_l$ being the
upper and lower boundaries) at the location of
the innermost field maximum,
the number of wavelengths in radial direction, $n_R,$  specified
in order to define the net and zero net flux fields, respectively,
the location of the boundaries to the three different
regimes of the R-domain, $R_1,R_2$ and $R_3$,
the extent of the domain in $\phi$ and z, and the
duration of the run in units of the inverse Keplerian angular frequency 
at the inner  boundary located at $R_0\equiv 1.$ 
Typically the runs last for
about 20 orbits measured at the
outer radius of the active Keplerian region located at 
$R  \sim 4$, although we  performed one long run  for a larger
disk with $R_2/R_0 = 7.2$ for a time exceeding 50 orbits
at the outer boundary of the Keplerian domain in order
to confirm the persistence of the turbulence up to such times.
For computational purposes,
the unit of length is taken to be
the inner boundary  radius, thus  $R_0 = 1,$
the unit of mass is the central mass, and the unit of time
is the inverse Keplerian angular frequency at the inner boundary  radius, 
or the period  there divided by  $2\pi.$

\subsection{Numerical procedure} \label{S2}
\noindent
The numerical procedure
follows the method of characteristics
constrained transport MOCCT as outlined in Hawley \& Stone(1995)
and implemented in the ZEUS code.
Alfv{\`e}n  wave characteristics are used to integrate
the induction equation
and to evaluate the Lorentz force.
The evolution of the magnetic field, ${\bf B}$, is constrained to enforce
$\nabla \cdot {\bf B} =0$ to machine accuracy.
The code has been developed from  a version
of NIRVANA originally written by U. Ziegler
(see Ziegler \& R\"udiger 2000 and
references therein), and
has been previously used for 2D hydrodynamic simulations
of viscous disks interacting with migrating Jovian mass planets.
The results have been validated by detailed comparison with other codes
and independently obtained results (Bryden et al. 1999, Nelson et al. 2000,
Kley, Angelo \& Henning 2001).
Shearing box MHD simulations have also been  performed by
 Ziegler \& R\"udiger (2000).

\noindent
The time step is limited by the Courant condition.
In addition, we have implemented the
numerical technique used by Miller \& Stone (2000) to  prevent
the time step becoming too small
by not allowing the Alfv{\`e}n speed to  exceed a  limiting 
value in regions of very low
density. Tests have shown that this makes no significant difference
to the results obtained, 
while circumventing the severe drop in the time step which
would sometimes make practical continuation
of simulations impossible.
We also set a global floor for the density.
On performing a series of exploratory runs with different values
for the limiting speed and the density floor, 
we found that simultaneous application of the two methods
leads to the best results in a most economical manner.
However, as stated above, the results were unaffected by reasonable
changes to the floor or the limiting velocity.
For the runs listed in table 1,
the density floor was $\rho_{fl}=10^{-3}\rho_0$ and the limiting
velocity was $v_{lim}=0.3$.

\begin{table}
\begin{center}
\begin{tabular}{cccccccccccc}
 \hline
       &       &          &                                &                                 &       &     &      &       &        &   &     \\ 
 run &  B    &  $\langle \beta \rangle $ & $N_z\times N_R\times N_{\phi}$ & $\frac{z_u-z_l}{\lambda_{max}}$ & $n_{R}$&$R_1$& $R_2$& $R_3$ & $\phi$ & z & time\\
    &    &    &      &
               &       &     &   &       &        &   &      \\ 
 \hline
b1&0-net $B_z$& 105   & $36\times 334\times 108$&2  &6 &1.2 & 3.7  &4.5  & $\pi/2$  &$\pm$0.2& 526\\
b2 & ---        &316 &  $36\times 334\times 108$&4  & ---&--- & --- &---   & ---  &--- & 592 \\
b3& ---  &1005       &  $36\times 334\times 108$&6 & ---&--- & --- &---   & ---  &--- & 697 \\
b4& --- &143 &   $36\times 167 \times 108$ &0.5 & 3 &1.5 & --- &--- & --- &--- & 710 \\
b5& 0-net$B_{\phi}$ &32 &   $36\times 167\times 108$  &0.3 & ---&--- & --- &---   & ---      &---      & 739 \\
b6& 0-net $B_z$      &568 & $40\times 370\times 100$ &1 & 9&1.2 & 7.2 &8.8   & ---      &---      &6627 \\
b7& ---  & 563 & $60\times 370\times 100$ &2 & ---&--- & --- &---   & $\pi/3$  & $\pm$ 0.3    &  2818 \\
b8& ---  & 316& $54\times 334\times 108$  &1   &3   &--- & 3.7 &4.5  & $\pi/2$  &$\pm$0.2 &740 \\
b9& 0-net$B_{\phi}$ & 38 & $56\times 334\times 108$  &2   &---   &--- & --- &---   & ---  & ---     &741\\
b10& --- & 40 & $36\times 334\times 108$ &2  &---   &--- & --- &---   & ---& ---     & 750 \\
b11& 0-net $B_z$&80    & $34\times 132\times 34$ &3   & ---  &--- & 3   &3.6
  & $\pi/3$  &$\pm$0.18&1193 \\
b12& ---&438    & $54\times 132\times 64$ &4   & 6  &--- & ---  &---
  & ---  &$\pm$0.2&1381 \\

 \hline
n1& net $B_z$     & 120 & $44\times 202\times 54$  & 3   &0.5 & --- & 4   & 4.8 &  ---   &  $\pm$ 0.2  & 1444  \\
n2& ---           & 360 & $34\times 177\times 72$  & --- &--- & --- & 3.7 & 4.5 &  ---   &  $\pm$ 0.18 &  942  \\
n3& ---           & 120 & $44\times 202\times 102$ & --- &--- & --- & 4   & 4.8 & $\pi$  &  $\pm$ 0.2  & 628  \\
n4& net $B_{\phi}$& 7.6 & $34\times 132\times 34$  & 1   &--- & --- & --- & --- & $\pi/3$&   ---       & 1444 \\
 \hline
\end{tabular}
\end{center}
\caption{ \label{table1}  Parameters 
associated with the simulations
discussed in this paper. 
The first column gives the simulation label,
the second the nature of the 
initial field, the third the initial  value of 
$\langle \beta \rangle ={\int P d\tau \over \int {\bf B}^2/(8\pi) d\tau},$ with
the integrals being taken over the Keplerian domain,
the fourth the computational grid  and the fifth gives the number  
of wavelengths in the vertical domain of the most unstable MRI mode
calculated for the initial field using equation (\ref{lmax}).
The sixth
column gives the number of radial wavelengths in the initial
field, and the remaining columns give the boundaries and extents
of the radial, azimuthal and vertical computational domains
as well as the run time.
Models b1, b2, b3, b8, b9,
and b10 were all run with heating and cooling exactly in balance
while the remainder  had an isothermal equation of state 
with sound speed a fixed function of radius.}
\end{table}
\noindent
In all cases, the instability was initiated by applying a sinusoidal
perturbation in the radial velocity with an amplitude of $0.01c_0.$ 
The number of  wavelengths in the vertical domain  of the  most unstable MRI
mode calculated for the initial field 
at the location
of the innermost
 field maximum varies between 1 and 6  as specified in column 5 of table 1. The wavelength
 associated with the mode of maximum growth  was calculated  from 
\begin{equation}
\lambda_{max}=\sqrt{\frac{16\pi}{15}} \frac{B}{\sqrt{\rho} \Omega}.
\label{lmax}\end{equation}
We note that
due to the radial dependences of the density, angular velocity and
magnetic field,  $ \lambda_{max}$ is a function of radius.
 If $ \lambda_{max}$  is smaller than the extent of the z domain,
in the case of vertical fields,
 regions of stability may alternate with unstable regions.

\section{Angular momentum transport}\label{S3}
\noindent
In order to describe the behavior
of  the different models, it is helpful to use quantities that are
vertically and azimuthally averaged over the $(\phi, z)$ domain.
These are defined with an overbar such that for any quantity
$Q$
\begin{equation}
{\overline {Q(R,t)}} ={\int \rho  Q dz d\phi \over \int  \rho dz d\phi}.
\end{equation}                                
Note that although the numerical simulations
are done over a fraction of the full $\phi$ domain, because of periodicity
they can be stacked end to end so that without loss
of generality we can assume they occupy the full $2\pi$ when performing
azimuthal averages.

\noindent We also introduce the surface density 
\begin{equation}
\Sigma = {1\over 2\pi}\int \rho dz d\phi . 
\end{equation}

\noindent
Using the above quantities it is possible to describe the angular
momentum transport in the disk using functions dependent 
only on $R.$ In this way a connection to classical
viscous disk $'\alpha'$ theory can be made ( Balbus \& Papaloizou 1999).

\noindent We monitor the  vertically and azimuthally
averaged Maxwell and
Reynolds stresses, which are  respectively defined as follows:
\begin{equation}
T_M(R,t)=2\pi
\Sigma{\overline{\left({B_R(z,R,\phi,t) B_\phi(z,R,\phi,t) \over 4\pi\rho}\right)}}
\end{equation}
and
\begin{equation}
T_{Re}(R,t)=2\pi
\Sigma
{\overline{\delta v_R(z,R,\phi,t)\delta v_\phi(z,R,\phi,t)}}
\end{equation}
The velocity fluctuations $\delta v_R$ and $\delta v_\phi$
are defined through,
\begin{equation}
\delta v_R(z,R,\phi,t)=v_R(z,R,\phi,t)-{\overline{v_R}}(R,t),
\end{equation}
\begin{equation}
\delta v_\phi(z,R,\phi,t)=v_\phi(z,R,\phi,t)- {\overline{v_{\phi}}}(R,t).
\end{equation}
With these definitions, the Shakura  \& Sunyaev (1973) 
$\alpha$ parameter for the
total stress is given by
\begin{equation}
\alpha(R,t)=\frac{T_{Re}-T_M}{2\pi
\Sigma{\overline{ \left(P/\rho\right)}}},
\end{equation}
where the pressure average is taken at the actual time rather than
at $t=0$ as is sometimes found in the literature. 
With the above definitions the vertically and azimuthally averaged
continuity equation (1) may be written
\begin{equation}
\frac{\partial \Sigma}{\partial t}+\frac{1}{R}
\frac{\partial\left(R\Sigma\overline{v_R}\right)}{\partial R}=0 \label{cona}
\end{equation}                       
and the  vertically and azimuthally averaged azimuthal
component of the equation of motion(2) can be written in the form
\begin{equation}
\frac{\partial \left(\Sigma \overline{j}\right)}{\partial t}
+\frac{1}{R}\left(
\frac{\partial\left( R\Sigma\overline{v_R}\overline{j}\right)}{\partial R}
+\frac{\partial\left(\Sigma R^2\alpha\overline{P /\rho}\right)}{\partial R}
\right) =0.\label{mota}
\end{equation}                 
Here $j=rv_{\phi}$ is the specific angular momentum.
Equations (20) and (21)
are identical  to what is obtained in viscous $'\alpha'$ disk theory
(Balbus \& Papaloizou 1999). However, as noted by those authors
significant differences may occur when the energy balance is considered.
From equations (1) and (2)
we may derive the rate of energy dissipation, $\epsilon_{\nu}$
 and doing $PdV$ work per unit volume as
\begin{equation}
\frac{\partial (\rho\epsilon)}{\partial t}+ \nabla \cdot {\bf F}
=-P\nabla\cdot{\bf v}-\epsilon_{\nu}=-Q_T,
\label{energ}\end{equation}   
where, the energy  per unit mass and flux are respectively
$$\epsilon={1/2}{\bf{v}}^2 + \Phi + {\bf B}^2/(8\pi \rho) $$
and 
$${\bf F} =\rho
{\bf{v}}\left({1/2}{\bf{v}}^2 + \Phi +\frac{{\bf B}^2}{4\pi \rho} +{P\over\rho}\right)
-\frac{({\bf{v}}\cdot{\bf{B}}){\bf{B}}}{4\pi}.$$
To reorganize the above into a form more related to
$'\alpha'$ disk theory we define   an energy  per unit mass $\epsilon_k(R)$
and specific angular momentum $j_k(R).$ These can be used to define
an angular  velocity $$\Omega_k= 
\frac{{d\epsilon_k(R)\over dR}}{ {d  j_{k}(R)\over dR}},$$
and may be chosen to correspond to a  free particle in circular
Keplerian orbit.
The latter assumption is not necessary and any convenient
values could be adopted in principle. Note that in the  general
non Keplerian case
$\Omega_k \ne j_k/R^2=v_k/R.$ Performing an azimuthal and vertical average
on equation(\ref{energ}) and subtracting (\ref{mota}) after 
multiplying by $\Omega_k,$ we may write the energy balance
in an alternative form
\begin{equation}
\frac{\partial(\Sigma \bar{\cal{E}})}{\partial t}+ 
\frac{1}{R} \frac{\partial(R  \bar {\cal{F}})}{\partial R}
+\Sigma\left(\bar{v_R}(\bar{v_{\phi}}-v_k)
+\alpha(R){\overline{ P/\rho} }\right)R\frac{d\Omega_k}{dR} =
-{1\over 2\pi}\int Q_Tdzd\phi.
\label{energbal}\end{equation}       
Here $${\cal{E}}= \epsilon-\epsilon_k -\Omega_k(\bar{j}-j_k)$$ and
$${\cal{F}}=\Sigma v_R\left(\epsilon-\epsilon_k -\Omega_k(\bar{j}-j_k)
+{P\over\rho}+\frac{{\bf B}^2}{8\pi\rho}
- R\Omega_k{\bf{\hat{\phi}}}\cdot({\bf{v}}-\bar{{\bf{v}}})\right)$$
$$-\Sigma({\bf{v}}-R\Omega_k{\bf{\hat{\phi}}})
\cdot\frac{{\bf {B}}B_{R}}{4\pi\rho}$$
It can be argued (adopting Keplerian values for
$\epsilon_k, j_k$)  that in a thin disk that is in a time average near
 steady state
the term proportional to $\alpha$ on the left hand side 
 of (\ref{energbal})  is second order
in $c/v_{\phi},$ while the others are at least third order
(Balbus \& Papaloizou 1999). To do this one needs to
extend the azimuthal and vertical
average to incorporate an additional
time average which is to be carried out over a time long
compared to the orbital period but
short compared to a supposed much longer evolutionary time scale.
The relation of 
$\alpha$ to energy dissipation that
holds in standard $'\alpha'$ disk theory is then recovered
in the thin disk limit. 

\noindent
However, in some practical cases, including the calculations
reported here, which have
somewhat large values of 
$c/v_{\phi}\sim 0.1,$ important deviations may occur.
For example the  ratio of the  contribution of the
pressure flux term $\bar{ v}_R P$ to the term 
 $\propto \alpha$ in (\ref{energbal})
may be estimated as $\bar{v}_R/(\alpha v_{\phi}).$
In our calculations, at any particular time,
this can be of order unity indicating that
energy redistribution through pressure wave modes may be important
and signaling  departures from standard viscous disk theory
as far as the energetics is concerned. 
Thus the adopted $'\alpha'$ parameterization may be useful
only as far as angular momentum redistribution  is concerned.
In this context in distinction to standard viscous disk theory,
there is no reason why $\alpha$ should be invariably  positive.
Positive definite dissipation may still occur provided
the $\cal{F}$ flux terms provide 
a source (see also Balbus \& Papaloizou 1999).
In general the magnetic stresses always give a  positive
contribution to $\alpha.$ However, the contribution of the
Reynolds stress can be strongly variable in space, time 
and sign.

\noindent
We  monitor the average time dependent (radial) Mach number,
\begin{equation}
M_s(R,t)=\frac{\bar{v_R}}{  c(R)}.
\end{equation}
In general $\bar{v_R}$ exceeds
the expected mean viscous inflow velocity 
\begin{equation} v_{\nu}=
\alpha c^2/\bar{v_{\phi}}\label{viscv} \end{equation}
\noindent appropriate to a Keplerian disk because
there are fluctuations
in the average radial velocity with near zero  mean (compared
to their amplitude).

\noindent
Following Hawley(2000) we use the vertically and azimuthally averaged
stresses as a measure of angular momentum transport through
$\alpha(R,t)$ defined above.
As indicated above one can incorporate a time average
with a view to indicating the behavior of a 
disk evolving on a time scale
long compared to the orbital period.
The form of the
averaged equations given above  is unchanged but 
the mean radial inflow velocity
is significantly reduced,
becoming comparable to the viscous inflow speed given by equation
(\ref{viscv}).

\noindent
In addition, we have considered mean values
of $\alpha$ which are obtained  from  volume
averaging over the Keplerian
domain.
Thus we denote the volume averaged value of $\alpha,$
as a function of $t$,
through

\begin{equation}
\langle \alpha (t)\rangle  = 
{\int \limits^{R_2}_{R_1}\alpha(R,t)RdR\over (1/2)(R_2^2 - R_1^2)}. 
\end{equation}
\noindent
In practice we find for runs with zero net flux which do not show
extreme density contrasts that radial averaging of the total stress
over the Keplerian domain (I) 
gives a value with much reduced temporal fluctuations
compared to those  occurring at a particular point  once the
turbulence is established (e.g. the  volume average is never negative).
Most of the fluctuations seen  in
our runs appear
to be due to the 
Reynolds stress.  Once a time average is performed the contribution
of the latter to the total stress is relatively small $\sim 1/3$
in line with local simulations (e.g. HGB96).

\noindent
Power spectra of the  vertically averaged
magnetic field components are calculated as
the squared Fourier amplitude of an azimuthal mode $m$
as a function of radius as follows
\begin{equation}
\mid a_{mbi}(R,t)\mid^2=\left| \int\limits^{z_u}_{z_l}\int\limits^{2\pi}_0B_i(z,R,\phi,t)
e^{-im\phi}d\phi dz\right|^2, \label{spectra} 
\end{equation}
where $i=z,R,\phi$.

\section{Simulations with zero net flux fields}\label{S4}
\noindent
Provided there is  zero field  at the radial  boundaries, in 
a global simulation of the type considered here,  both the total vertical
and the total toroidal flux threading the system are conserved.
The latter is guaranteed by the periodic boundary conditions
in the vertical direction.
As long as there is no flux entry through the radial boundaries 
simulations can therefore be characterized by the amount of net flux
they contain. In particular, if there are no other conserved quantities
that can distinguish different simulations, one might expect 
that (allowing for numerical limitations)  all simulations
with zero  initial net flux should approach the same turbulent state.
This has been found to be the case for shearing box simulations
(HGB96). However, the shearing box set up
is special in that fully periodic
boundary conditions in shearing coordinates
guarantee  no flux entry into the system for all time.
In addition, one can search for a time averaged steady state over
arbitrary time intervals.
\noindent
The set up of the global simulations considered here only allows
for testing that simulations starting  with zero net flux
approach the same steady state in a more restricted sense.  
Although we expect magnetic energy to  grow through the action of instabilities
only in the Keplerian domain, we inevitably find diffusion of field into the
inner boundary layer region as well as the outer stable 
 region.
When there is significant shear in the boundary layer region,
we find the field can grow through the winding up of the poloidal 
field to produce a toroidal field with significant magnetic energy.
However, this phenomenon may be considerably delayed by starting a simulation
with initial field set to be non zero only at large distances from the
boundary regions. In this respect 
we remark that toroidal fields are preferred to vertical
fields and the absence of shear in the inner layer delays the build
up of strong fields there.
In any case there is no guarantee of conservation
of magnetic flux in the 
Keplerian domain once   there has been significant  field
diffusion out of that region.
Accordingly, checking whether solutions approach the
same turbulent state in the Keplerian domain 
has been limited to the     
situation before such diffusion has occurred.
If convergence of the solutions occurs, then it should of course
hold at a later stage when interaction with  the boundary  regions occurs.
In this   context we have found that different prescriptions for
the inner boundary layer (i.e.\ $n_b=1$ or $n_b=3$) do not seem to
affect the situation in the Keplerian domain even
though the  behavior in the  inner boundary  layer region 
may be significantly different.
\noindent
In this section we investigate the disk response to varying
initial fields with zero net flux in order to establish
that when  these are of small enough radial scale
and adequate amplitude essentially the same state results.
Other effects such as those of the numerical resolution,
disk aspect ratio and size are also explored.
\noindent
We investigate  both initially prescribed  vertical
and toroidal magnetic fields.
We find essentially the same final state when 
the field has a small scale
compared to the current radius and a local 
quasi-steady turbulent state can be achieved on a time 
scale short compared to the
global evolution time of the entire disk. 
Initial large scale fields with significant 
magnetic energy may not lead to the same state 
as small scale fields on a time scale
short compared to the evolution time scale of the disk 
or on one that can
be reasonably followed here.

\subsection{ Disks with initially vertical fields}
\begin{figure}
\centerline{
\epsfig{file=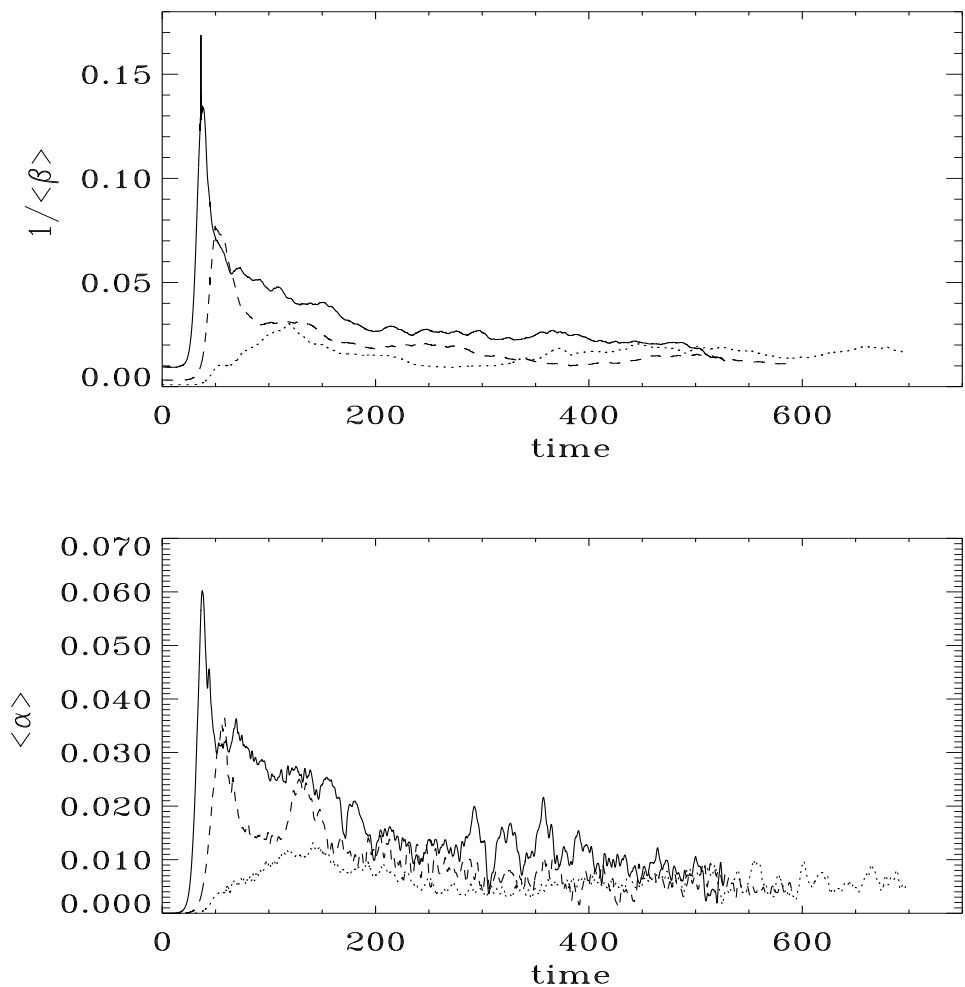,height=14.cm,width=14.cm,angle =360} }
\caption[]{ Magnetic energy ($1/\langle \beta \rangle)$ in the Keplerian domain
expressed in units of the volume integrated pressure
as a function of time
for the three runs b1, b2, b3 (upper panel).  
Run b1 corresponds to
the solid curve and b3 to the dotted
curve. In spite of very different initial
behavior for $0 < t < 200$ among these cases they all approach
a situation where in a time average
sense  $1/\langle \beta \rangle \sim 0.01.$ 
The value of $\alpha$
volume averaged  over
the Keplerian domain
$\langle \alpha \rangle$ is plotted in the lower panel as a function of time.
These, as many other cases, eventually
attain $\langle \alpha \rangle \sim 0.005.$}
\label{fig1}
\end{figure}

\noindent
We first consider the three runs b1, b2, b3 that
start from an initial vertical field defined between 
$R_b = 4/3$ and $R_{b1} = 10/3.$ In all of these cases
$n_b =3$ and $n_c=0.625$ in region II. The runs differ only in the amplitude
of the initial vertical field such that the initial $\langle \beta \rangle$
varies by an order of magnitude ranging between $\sim 100$
and $\sim 1000$ (see table 1).
Many features characteristic for simulations b1, b2, and b3 are also found in others
described below. 
The magnetic energy in the Keplerian domain
expressed in units of the volume integrated pressure
(or $1/\langle \beta \rangle$) is plotted as a function of time 
for the three runs in figure \ref{fig1}.

\noindent
As found with  simulations performed in a shearing box
(Hawley et al. 1995, Stone et al. 1996, 
Brandenburg et al. 1995, Ziegler \& R\"udiger 2000)
and existing global simulations (Hawley 2000, Krolik \& Hawley 2000),
the simulations with smaller initial 
$\langle \beta \rangle$ 
show the development and growth of  channel solutions.
This is manifest in the early evolution of b1  through
the strong peak in $1/\langle \beta \rangle$ that occurs at $t \sim 30.$
Streams of fluid at different vertical levels,
alternately moving in opposite radial  directions,  evolve and persist
over a period of several orbits at the corresponding radius. 
This phase is associated with large values of $\alpha$
peaking at 0.3-0.6 at smaller radii, where the instability grows first
due to the smaller rotation period.
This behavior is barely noticeable in run  b3. 
We comment that there is
not yet a significant penetration of the  magnetic field into
the boundary regions during the time interval shown.
In spite of very different initial
behavior for $0 < t < 200$ among these cases they all  approach
 $1/\langle \beta \rangle \sim 0.01$ (in an average sense) as in many
other cases we have run (see below). 
After $ t \sim 200,$   turbulence  is established
and maintained throughout the disk.
\noindent
At this stage, it is found that 
$\alpha$ exhibits strong variations in time and with radial
distance, sometimes up to one order of magnitude.
The value of $\alpha$
volume averaged  over
the Keplerian domain
$\langle \alpha \rangle$ is plotted as a function of time
in the lower panel of figure \ref{fig1}
for the three runs.
The volume averaged $\alpha$ shows strong fluctuations
on all time scale, which are, however, less than those found
at any particular radius.  
Closer inspection reveals that most of the fluctuation
is due to the Reynolds stress which may contribute
up to one half of the total. The Maxwell stress is found
to be always positive and to vary  less strongly.

\begin{figure}
\centerline{
\epsfig{file=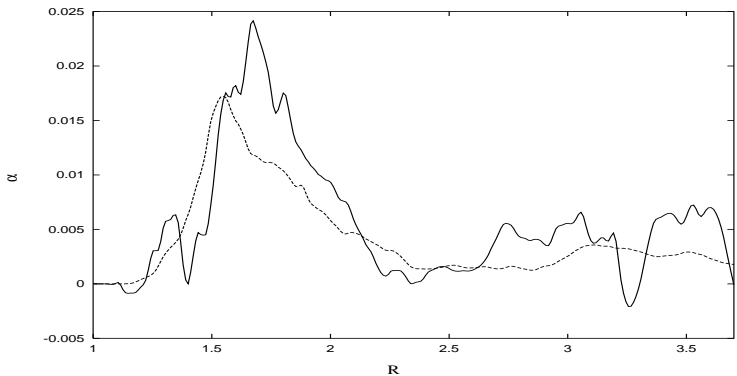,height=10.cm,width=5.cm,angle =270} }
\caption[]{ In this figure  $\alpha(R,t)$ (solid curve)
is plotted
against dimensionless radius
for run $b3$ at  time 643. The dashed curve gives the
contribution of the Maxwell stress.}
\label{fig2}
\end{figure}
\noindent
Simulations b1, b2, b3
attain time averaged values  of
$\langle \alpha \rangle \sim 0.005 $ as in many
other cases we have run (see below). This value is similar
to what is
seen in
shearing box simulations
starting  with weak zero net flux fields
(Brandenburg et al. 1995, Hawley et al.  1995).
We also comment that we obtain a similar 
approximate correlation between
 $(R, \phi)$-stress and energy as found in the shearing box simulations
of HGB96. This was verified in all
simulations with zero net flux and  
can be expressed, after averaging out short term variations, as
$\langle \alpha \rangle \sim  0.5/\langle \beta \rangle.$

\begin{figure}
\centerline{
\epsfig{file=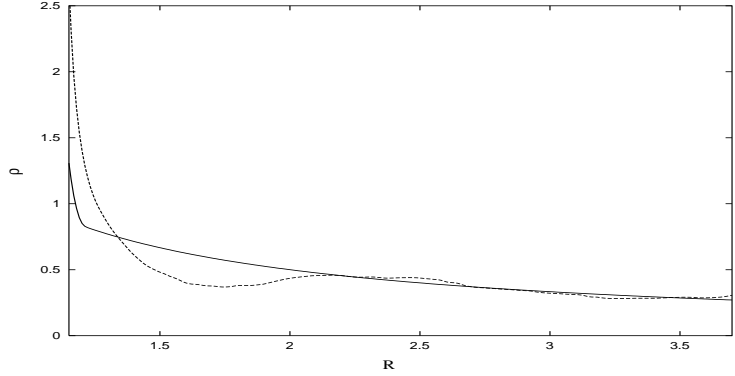,height=10.cm,width=5.cm,angle =270} }
\caption[]
{Radial dependence of the vertically and azimuthally averaged density
(given in arbitrary units)
for run b3
at time 643
(dashed line). The initial profile 
is given by the solid line. Note that
at the later time, the density depression
apparent between $R=1.5$ and  $R=2$ together with the
density increase at smaller radii is indicative of the
mass accretion process.}
\label{fig3}
\end{figure}
\begin{figure}
\centerline{
\epsfig{file=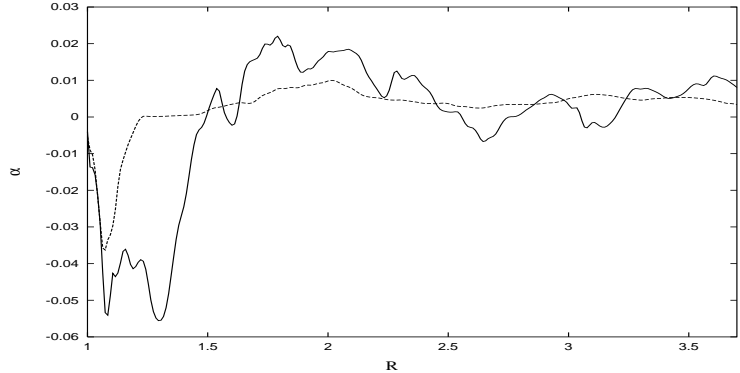,height=10.cm,width=5.cm,angle =270} }
\caption[]
{Radial dependence of $\alpha(R,t)$ (solid curve) 
for run $b1$ at time 540. There is significant
field penetration into the boundary layer
region as indicated by the negative values there.
The Maxwell stress  (dashed curve)
is also  negative in that region.}
\label{fig4}
\end{figure}
\noindent
We plot $\alpha(R,t)$
against radius
for run $b3$ at time 643 in figure \ref{fig2}.
No significant field has yet penetrated
into the boundary layer region as can be seen
by the small values there. Note that at this time 
and for this simulation
there is a strong peak at smaller radii.
However,
such a peak may occur in the center of the Keplerian domain
at other times and
in other simulations (see below).
The vertically and azimuthally averaged density
corresponding to figure \ref{fig2}
is plotted in figure \ref{fig3} together with the initial profile.
Comparison between the two curves shows evidence of accretion
with material moving into the inner region
from the outer parts of the disk. 

\noindent 
Although it can be delayed by starting with initial
data that is non zero far away from the inner boundary
layer, the field inevitably diffuses into this region 
with the degree of penetration increasing with time.
This is the case even though the shear indicates stability
with respect to the MRI.
The pattern of behavior is the same for all runs.
For runs b1, b2 and b3, which have $n_b=3$, the strong shear
causes the toroidal field to build up and the value
of $\alpha(R,t)$ to become negative corresponding to
inward angular momentum transport. 

\noindent
As an illustrative example
to indicate these points, we plot
$\alpha(R,t)$ 
against radius
for run $b1$ at  time 540 in figure \ref{fig4}.
The largest magnitudes of $\alpha \sim -0.05$
occur in this region which tends to expand outwards.
Nonetheless, radial motions remain subsonic. Eventually magnetic
contact with the inner boundary zones occurs causing a violation of the
conditions required for conservation of flux in the computational domain.
This effect may be delayed by choosing a more extended inner boundary region.

\subsection{ Disks with initially toroidal fields and
the effect of vertical resolution}
\noindent
We now describe models b8, b9 and b10.
We set $n_b=3, n_c=0.625$  in region II
and defined the magnetic field between $R_{b}=7/3$ and $R_{b1}= 10/3$
for b8 and between $R_{b}=4/3$ and $R_{b1}= 10/3$
for b9 and b10.
These runs were carried out to check that simulations
beginning with  small scale zero net toroidal fields
led to the same state as those starting from vertical fields.
For the purpose of comparing with a vertical field run, simulation b8 started with a poloidal field with the same initial
magnetic energy as b2, but with higher
vertical resolution.
Simulation b10 started with a toroidal field and had 
the same resolution as b1, b2 and b3, 
while b9 had higher vertical resolution.
Because of the very much weaker instability
apparent in the cases with an initial toroidal field,
these could be started with significantly larger
magnetic energy. No early
channel phase occurs.  In fact in these cases  the magnetic energy
decreases  due to  reconnection
of oppositely directed field lines in the initial phases of the MRI.
\begin{figure}
\centerline{
\epsfig{file=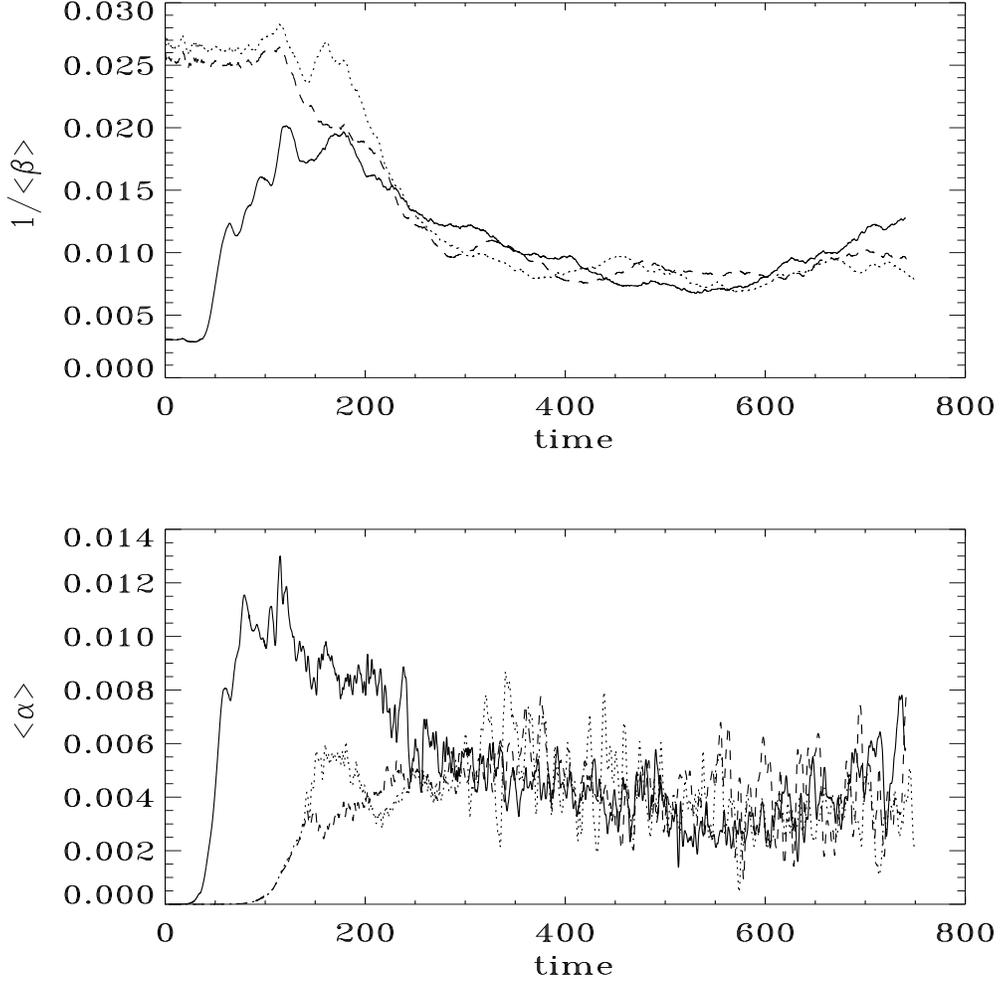,height=14.cm,width=14.cm,angle =360} }
\caption[]{Magnetic energy in the Keplerian domain
expressed in units of the volume integrated pressure,
as a function of time 
for the three runs b8 (solid curve), b9
 and  b10 (dotted curve) in the upper panel.  The  first
of these starts with a vertical field
while the other two, differing in vertical
resolution, start with a toroidal
field with significantly larger energy.
All these cases  eventually approach
 $1/\langle \beta \rangle \sim 0.01.$
In the lower panel the value of $\alpha$
volume averaged  over
the Keplerian domain
$\langle \alpha \rangle$ corresponding to these runs is plotted as a function of time
(b8, solid curve, b10, dotted curve).}
\label{fig5}
\end{figure}
\begin{figure}
\centerline{
\epsfig{file= 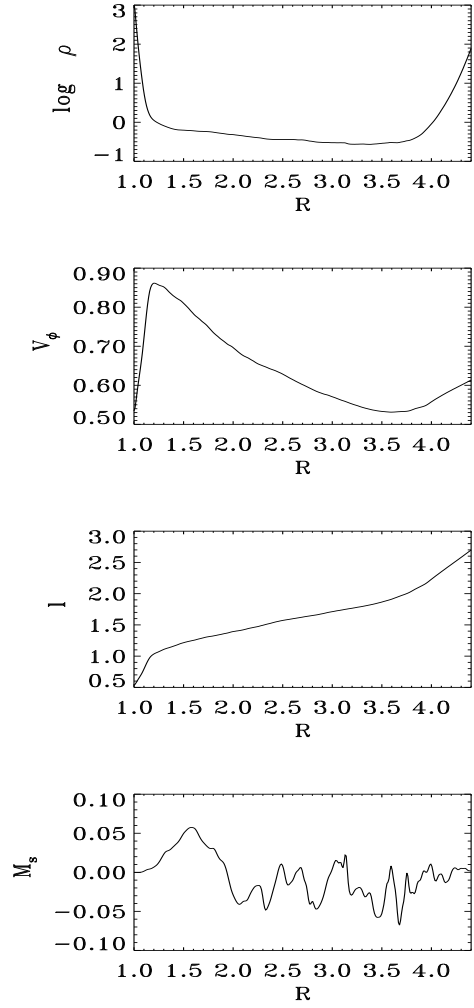,height=14.cm,width=7.cm,angle =360} }
\caption[]{From the top panel
down, vertically
and azimuthally averaged density,
azimuthal velocity, specific angular momentum
and radial Mach number near the end of simulation b8. }
\label{fig6}
\end{figure}

\noindent
The magnetic energy in the Keplerian domain ($1/\langle \beta \rangle)$
expressed in units of the volume integrated pressure
is plotted as a function of time
for these three runs in the upper panel of figure \ref{fig5}.  
There is no significant penetration
of magnetic field into the boundary regions
at this stage.
In spite of  the different initial conditions
and the initially weaker instability in the models starting
from a toroidal field,
they all eventually attain
  $1/\langle \beta \rangle \sim 0.01$,
as was found for simulations b1, b2 and b3.
The increased vertical resolution of simulations 
b8 and b9 appears to have little influence
on the results.

\noindent
The value of $\alpha$
volume averaged  over
the Keplerian domain
$\langle \alpha \rangle$ is plotted as a function of time
for the simulations b8, b9 and b10
in the lower panel of figure \ref{fig5}. These values behave similarly to
those found for simulations b1,b2 and b3, and become
indistinguishable at later times,
with the time averaged 
value of  $\langle \alpha \rangle = 0.004 \pm 0.002.$

\noindent
For simulation b8 we present in figure \ref{fig6} the
vertically averaged density,
azimuthal velocity and radial Mach number profiles near the
end of the run. These are similar to the initial profiles.

\begin{figure}
\centerline{
\epsfig{file= 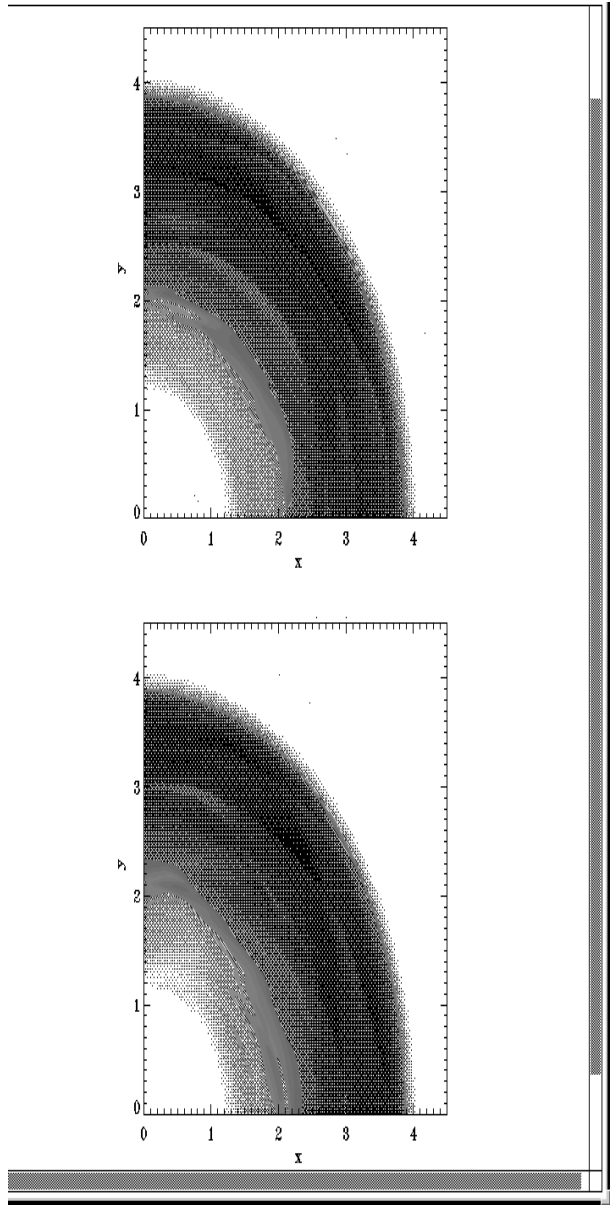,height=16.cm,width=8.cm,angle =360} }
\caption[]{Density contours in the mid plane
 near the end of the  runs b8 ( upper panel) and b9 (lowe panel). Although these runs
start from very different initial conditions,
the plots are similar.}
\label{fig8}
\end{figure}

\begin{figure}
\centerline{
\epsfig{file=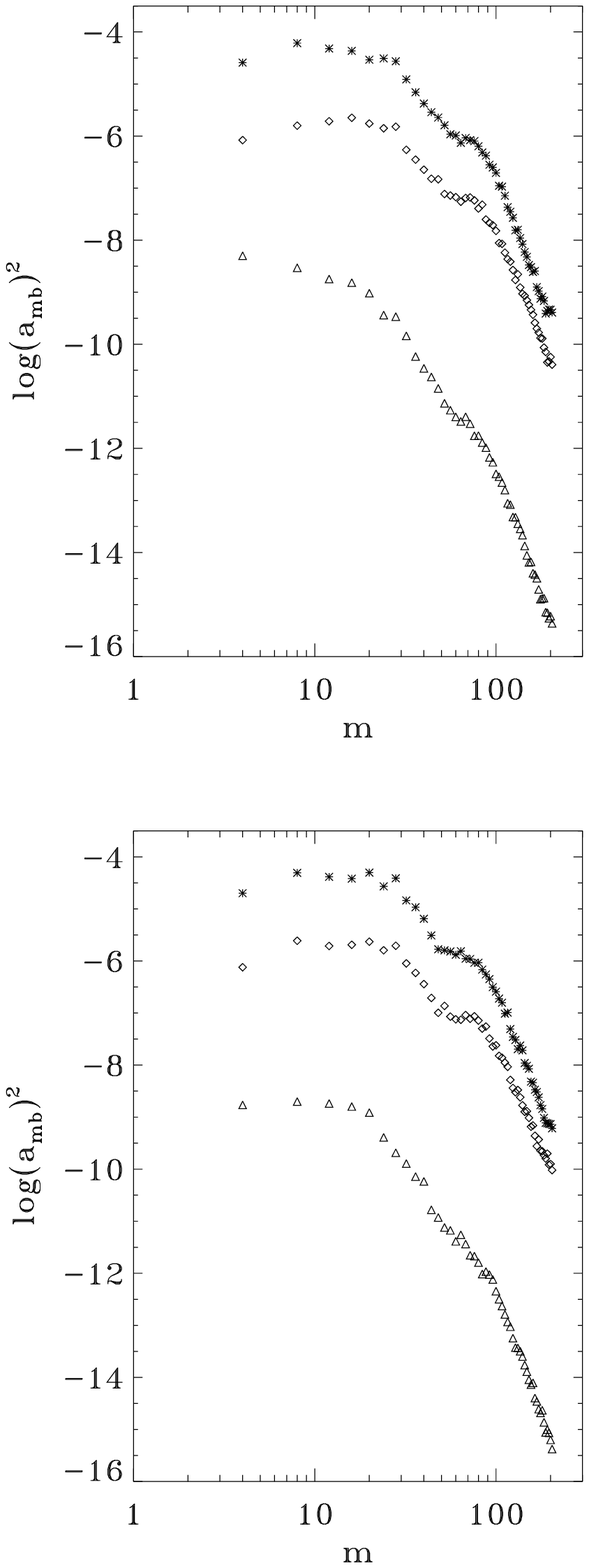,height=16.cm,width=7.cm,angle =360} }
\caption[]
{Azimuthal power spectra
(as defined in the text) of the vertically
integrated field components $B_{\phi}$ (asterisks), $B_R$ (diamonds) 
and $B_z$ (triangles) at $R=2$
for b8  (upper plot) and b9 near the end of their runs. Very similar
spectra occur at all radii in the Keplerian domain.
Although these runs
start from very different initial conditions,
the plots are similar.}
\label{fig9}
\end{figure}

\noindent
In order to emphasize the similarity between simulations with
initial
vertical and toroidal fields with zero net flux in figure \ref{fig8}
we plot the  mid plane density contours for simulations
b9 and b10 near the end of these runs.
Dark regions correspond to low density. Stochastic spiral
patterns are  visible.
In addition, we present the azimuthal power spectra for the
vertically averaged magnetic
field components calculated according to eq. (\ref{spectra})  at $R=2$ 
in figure \ref{fig9}. In fact these are characteristic of 
the turbulent state and are very similar at all radii.
There is a flat spectrum for small
azimuthal wavenumbers $m<10$ and a sharp cut-off at $m=10-20$ 
This behavior
is in agreement with the results previously obtained
by other authors (e.g. Armitage 1998,  Hawley et. al 1995).

\subsection{Large disks and long runs}
\noindent
We also ran two simulations, b6, and b7, for which the
disk was approximately twice as large as in the previous cases.
This has the consequence that the evolutionary  time scale
for the whole disk is about three times longer.
A larger disk enabled insertion of initial magnetic field
data further from the inner boundary region with
$R_b = 3.5$ and $R_{b1} = 6.5$.
In both these models 
$n_b=3$ and $n_c=0.625$ in region II.
 The resolution in simulation b6 was very similar
to the  cases discussed above. However, the
vertical domain extends over only one half
the disk  semi thickness in the outer part of the Keplerian domain.
In simulation b7, the azimuthal domain was contracted to
$\pi/3$ effectively increasing the azimuthal resolution by fifty percent.
In addition, the extent of the vertical domain was increased
by fifty percent with respect to b6, while maintaining the same resolution.

\begin{figure}[t]
\centerline{
\epsfig{file= 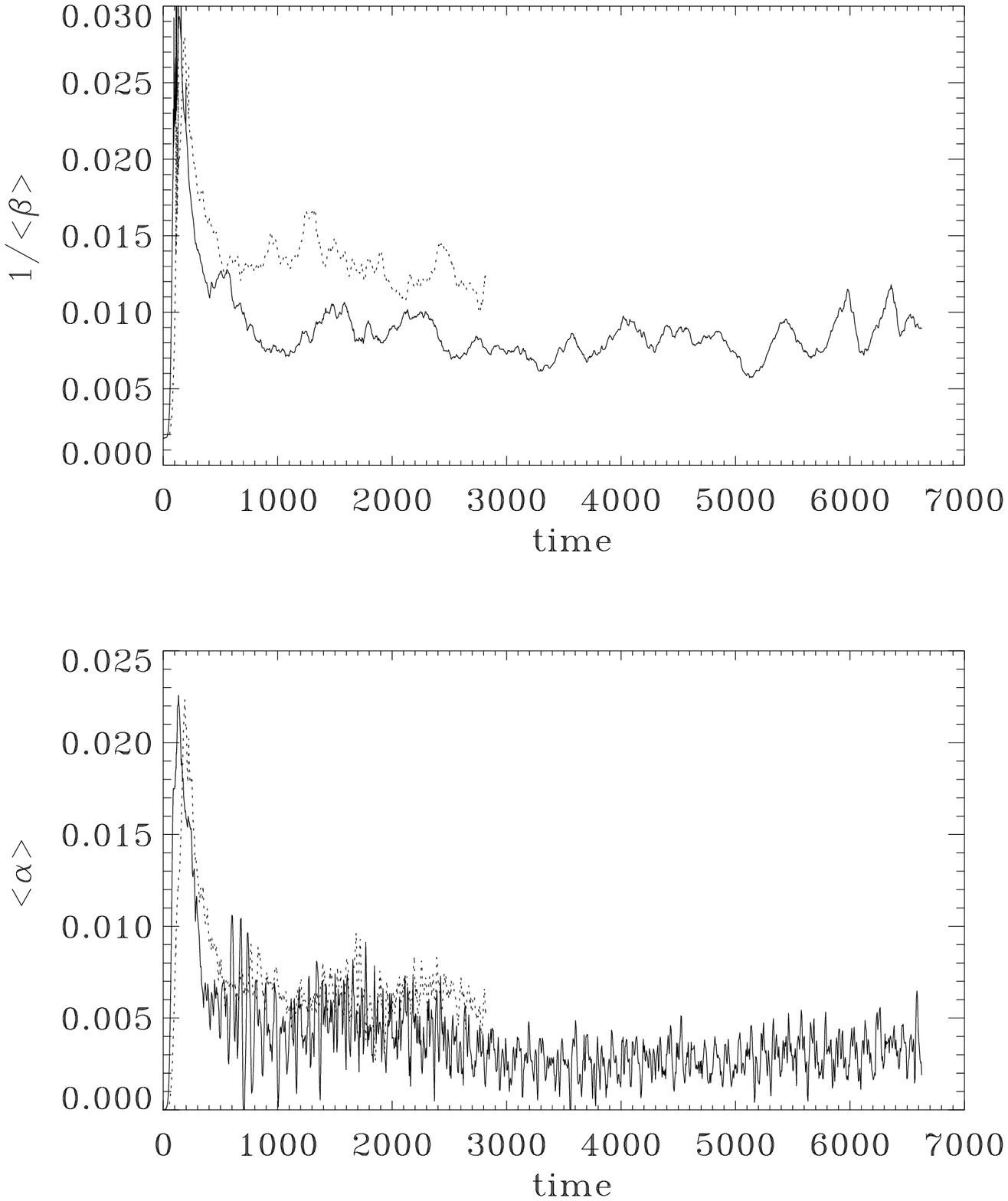,height=13.cm,width=14.cm,angle =360} }
\caption[]
{Magnetic energy in the Keplerian domain (in units of the volume integrated
pressure)
plotted as a function of time (upper panel)
for the  large disk runs b6 (solid curve) and b7.
In the lower panel we represented $\langle \alpha \rangle$
as a function of time
for these runs.}
\label{fig10}
\end{figure}

\noindent
The magnetic energy in the Keplerian domain
expressed in units of the volume integrated pressure
is plotted as a function of time 
for simulations b6 and b7 in the upper panel of figure \ref{fig10}.
These both
attain $1/(\langle \beta \rangle) \sim 0.01$
but with some indication of b7
tending to produce larger values than b6.
 A larger magnetic energy in the saturated turbulent state
in b7  might be expected from shearing box simulations ( HGB96).
These have indicated a dependence on numerical resolution
(Brandenburg et al. 1996), as well as
larger values for larger boxes. 
The effect of increasing the size
of the vertical domain in b7 is similar to increasing the size of a shearing box.
\noindent
The value of $\alpha$
volume averaged  over
the Keplerian domain
$\langle \alpha \rangle$ is plotted as a function of time
for  simulations b6 and b7 in the lower panel of figure \ref{fig10}.
These
attain $\langle \alpha \rangle \sim 0.004 \pm 0.002$ 
but with b7   tending to give  somewhat larger values in the mean. These
results are also consistent with the expected correlation
between  $(R,\phi)$-stress and magnetic energy mentioned above in the form
$\langle \alpha \rangle \sim  0.5/\langle \beta \rangle.$
Simulation b6 indicates survival of the saturated turbulent
state for more than fifty orbital periods at the outer
boundary of the Keplerian domain. At the end of simulation b7,
significant boundary layer penetration occurred producing negative $\alpha$
there. The situation then was very similar to that illustrated in figure
\ref{fig4} for  a smaller disk model.
Taken together, these simulations indicate
some  possible dependence of $\alpha$  on numerical resolution
and the extent of the vertical domain.

\subsection{A thicker disk}
\noindent
Simulations b4 and b5 
are characterized by larger semi thickness than the other runs,
with $c_0^2=0.04GM/R_0$, and  a wider uniformly rotating
inner boundary layer ($n_b=1, n_c=0.625$ in region II and $R_1 = 1.5).$ 
The initial fields were applied between $R_b = 7/3,$ and $R_{b1}=10/3.$
These runs were initiated with a zero net poloidal field (b4)
and a zero net toroidal field (b5). The initial magnetic energy
is about three times larger  in b5. These runs have lower absolute radial resolution
than those discussed above but  the same resolution per scale height.
Maintaining the extent of the vertical domain as in the 
previous runs b1, b2, b3 means that in these
two cases only one half
of the scale height is contained within the z-domain at the outer edge of the Keplerian domain.

\begin{figure}
\centerline{
\epsfig{file= 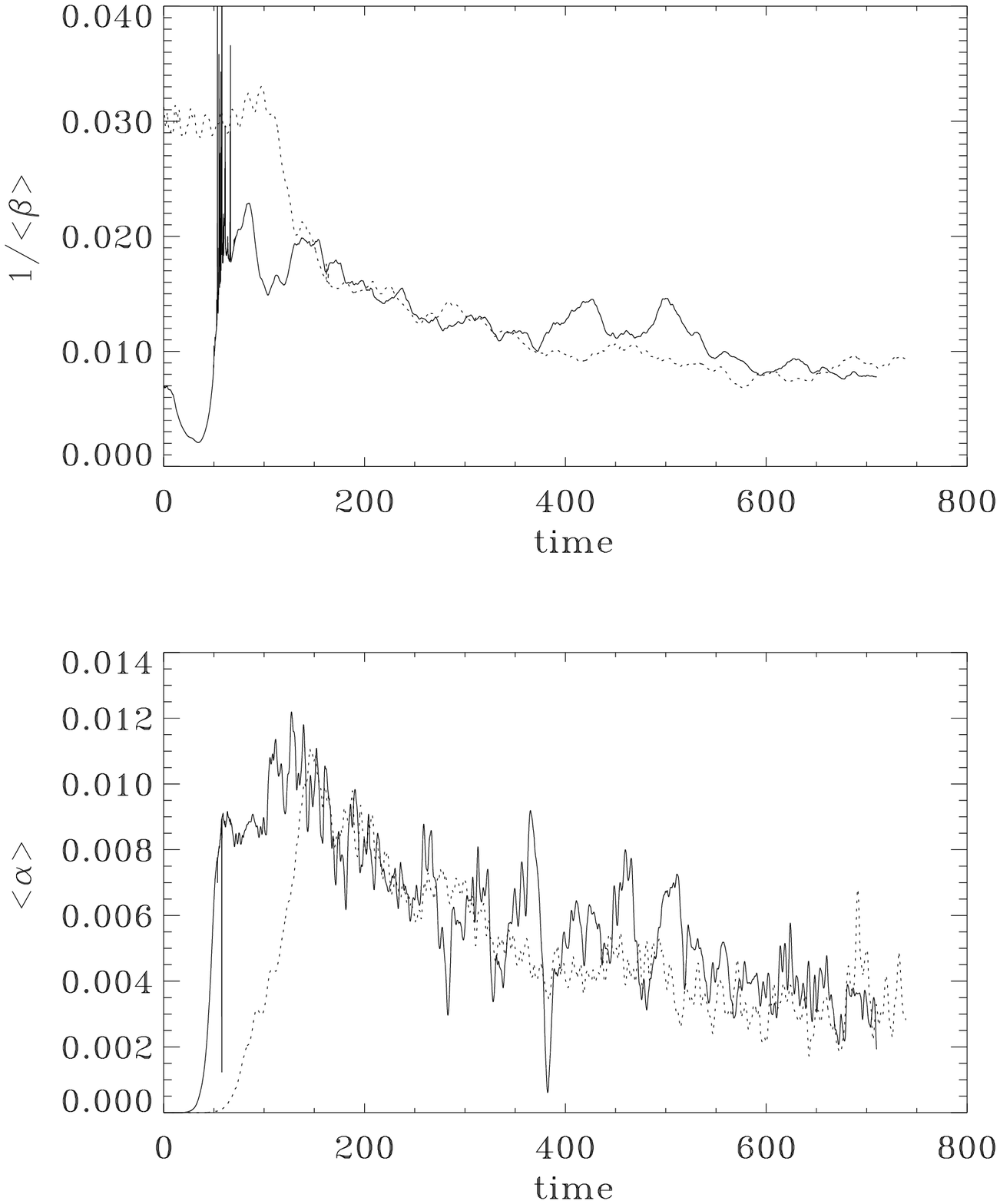,height=14.cm,width=14.cm,angle =360}}
\caption[]
{Magnetic energy in the Keplerian domain
expressed in units of the volume integrated pressure
as a function of time (upper panel)
for the  thick disk runs b4 (solid curve) and b5.
The corresponding value of $\alpha$
volume averaged  over
the Keplerian domain
$\langle \alpha \rangle$ as a function of time
is represented in the lower panel.}
\label{fig11}
\end{figure}
\noindent
The magnetic energy in the Keplerian domain,
$1/(\langle \beta \rangle)$, is plotted as a function of time
in the upper panel of figure \ref{fig11}.
In spite of the very different initial conditions these
simulations both approach  a state with $1/(\langle \beta \rangle)$
on average somewhat less than $0.01.$
The value of $\alpha$
volume averaged  over
the Keplerian domain
$\langle \alpha \rangle$ is plotted as a function of time
in the lower panel figure \ref{fig11}.  Similar values  to
those  obtained in the other simulations are found.

\begin{figure}
\centerline{
\epsfig{file=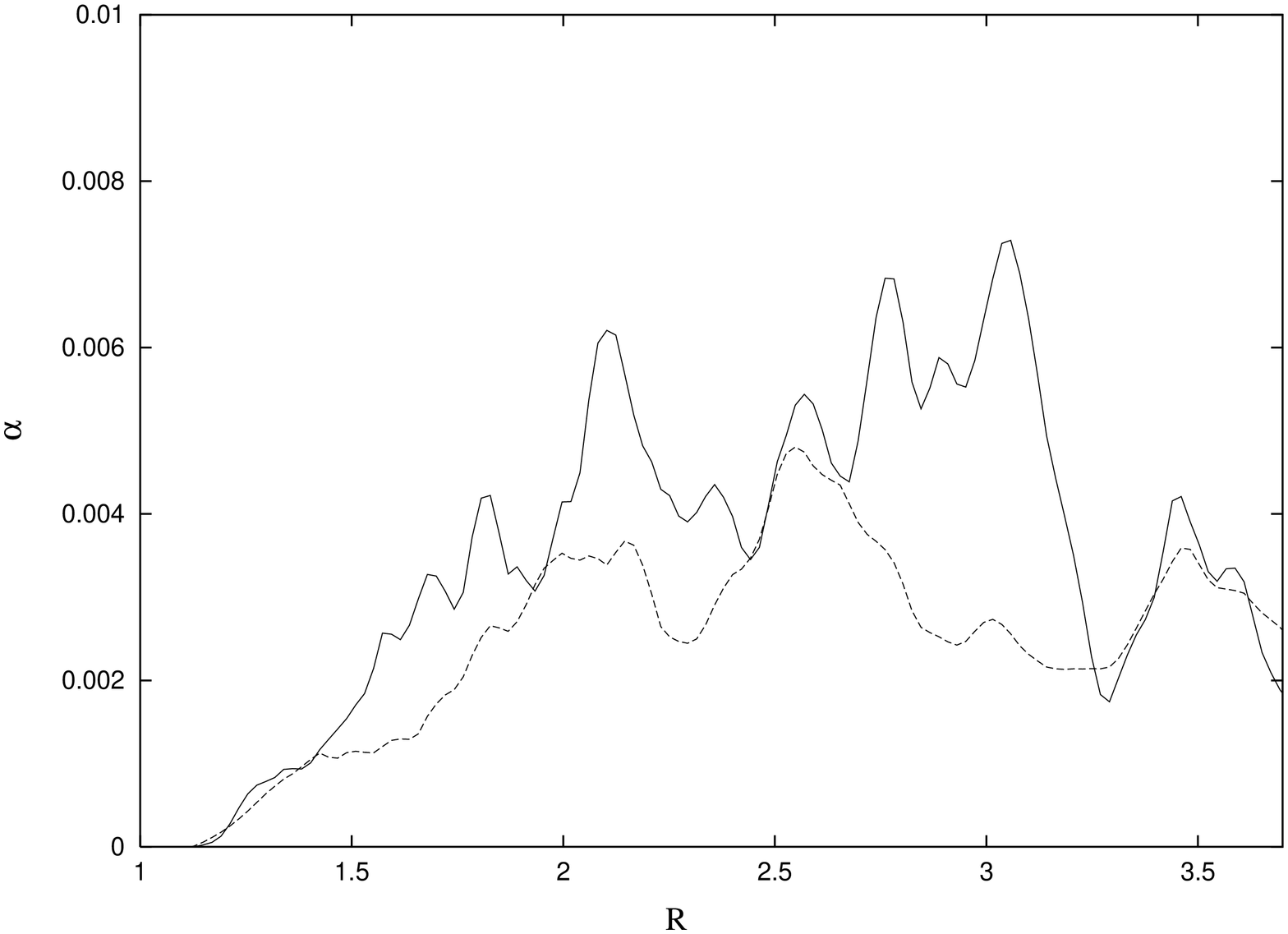,height=10.cm,width=5.cm,angle =270}}
\caption[]
{Radial dependence of $\alpha(R,t)$ (solid curve)
for run $b4$ at  time 590. The dashed
curve gives the contribution of the
Maxwell stress.}
\label{fig12}
\end{figure}
\noindent
We also plot  $\alpha(R,t)$ as a function of $R$ at time $t=590$
for simulation b4 in figure \ref{fig12}. This indicates a
peak in the  center of the Keplerian
domain. Even though there is some diffusion of magnetic energy
into the boundary layer at about 25 percent    
of the that in the Keplerian domain. There is little activity in the
boundary layer region while there is no contact
with the inner boundary zones, because of the low shear.
This has been found to be the case in these and other simulations. 
For the types of  computational set up used and the times we
have been able to run the  simulations,
we have found no evidence that the solution in the Keplerian domain
is significantly affected by the shear profile adopted
in the inner boundary layer. This is found to be the case (not illustrated here)
even when runs are continued well  beyond the point where magnetic contact
with the inner boundary zones occurs.

\subsection {Simulations with initial fields with large radial scale of
variation}
\noindent
We here describe simulation b11 which had an initial
vertical field with zero net flux with a larger
scale of radial variation (see table \ref{table1}) and $n_b=2$. The
field was initially applied between $R_b=1.32$ and $R_{b1}=2.76$.
This run was carried out with lower resolution than
the other simulations but in many ways led to similar results.
The large scale motion associated with the channel  solutions is completed 
after  $t \sim 240.$
After this time, as in other simulations, turbulence is established
and maintained throughout the disk.  
The largest values of
$\alpha(R,t) \sim 0.03 $ are typically
obtained close to the boundary layer and at the outer edge of
the Keplerian domain,
while in the middle of the active domain,
$\alpha(R,t) $ decreases to a mean value of 0.002. 
The large values of $\alpha$  
are connected with vertically and azimuthally
averaged density minima there, with a density contrast of up
to one order of magnitude with respect to the surroundings.
With this type of model,  unlike the previously described, 
convergence of different solutions, if it should occur,
is difficult to attain
on a reasonable time scale.

\noindent
As in other simulations, when the magnetic field penetrates
the boundary layer region  producing an
interaction between the boundary layer and the inner regions
of the  Keplerian disk, the boundary layer moves
to a somewhat larger radius
(from $R=R_1=1.2$ initially to $ R \sim 1.36$).
The  state reached after time 628
is therefore  somewhat
different from the initial state, with  surface density depressions
near the boundary layer and the outer
stable region. This behavior is typical for initial perturbations with
only one to a few wavelengths in the vertical domain 
for  the  most unstable MRI   mode and an initial field
with one to three maxima.

\subsection{Oscillations}
\noindent
The vertically and azimuthally
averaged radial 
Mach number remains subsonic throughout the simulations,
with the
typical radial velocity being $\sim 10$  times larger than the viscous
inflow velocity given by equation (\ref{viscv}).
It displays oscillatory  behavior  close to the outer stable boundary
region. Such oscillations, which occur in all models,
  were indicated  previously in  laminar  viscous
disk  modeling, away from the boundary layer, as a consequence of the viscous
overstability found by Kato (1978)
(e.g. PS). Here they arise as residuals after averaging turbulent
fluctuations vertically and azimuthally and have no obvious
connection to the earlier theories 
because the simple modeling used there 
does not incorporate the complication
that the time and length
scales of the oscillations and turbulence are not clearly
separable and so do not allow for a description using
an anomalous viscosity coefficient.
Such a description would only be expected to apply to
phenomena on a global length scale averaged over a  time scale
long compared to that characteristic of the turbulent  fluctuations.

\begin{figure}[t]
\centerline{
\epsfig{file= 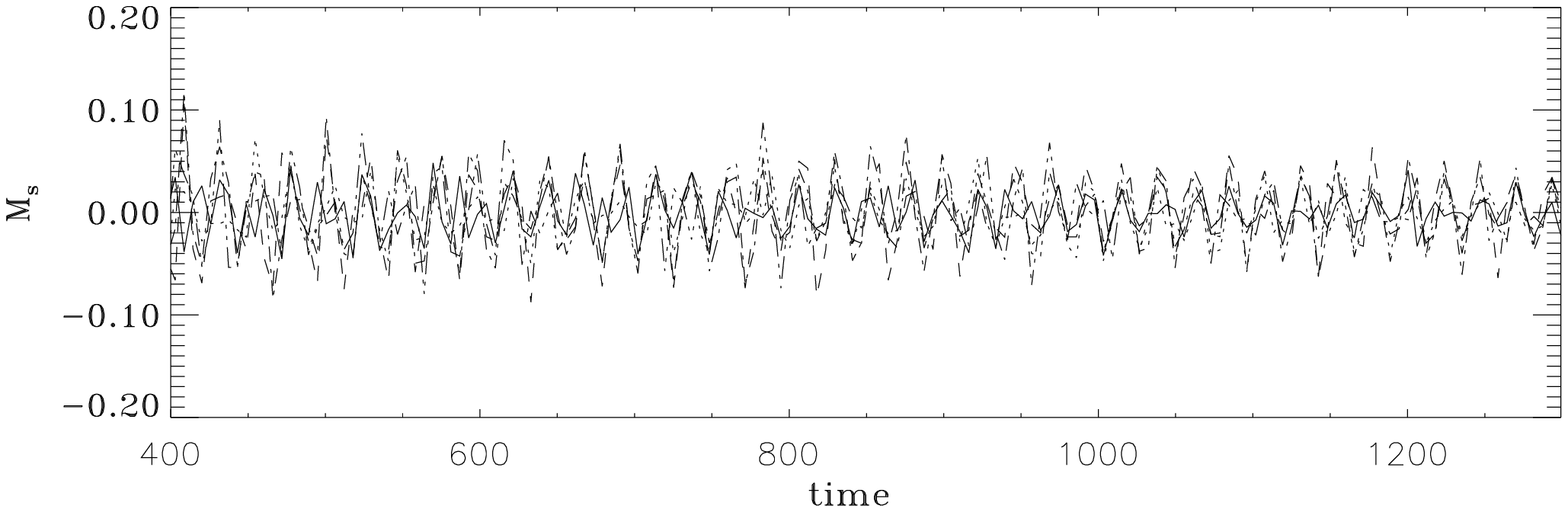,height=18.cm,width=7.cm,angle =90} }
\caption[]{Time evolution of the azimuthally and vertically averaged Mach number
at different radii, from R=2.04 to 3.04 in steps of 0.2 for run b12.}
\label{fig13}
\end{figure}

\noindent
In order to further
illustrate these oscillations, in figure \ref{fig13} we have represented the Mach number
as a function of time for simulation b12 which 
 had $n_b = 2 $ and 
as in all  previously discussed simulations
other than b11  was initiated with a small
scale field.
The  superposed curves apply to
locations in radius from $R=2.04$  to $R=3.04$ in steps of 0.2. The period of
the oscillations can be estimated with
some noise  to be  3.5 expressed in units of the Keplerian period at the inner
edge of the computational domain. The Keplerian period corresponding to this
range of radii ranges from 2.9 to 5.3. The smaller spread
seen in the oscillation periods
suggests that they are not entirely purely
local epicyclic oscillations
but 
that propagation of information from one location to another occurs in the simulations. 
A time series of the vertically and azimuthally averaged density
as a function of radius performed between two distinct times during
the simulation and covering several hundred time units
suggests  the presence of outward propagating disturbances
produced by the interaction
between the boundary layer and the Keplerian disk.
Some reflection from the  boundary of region III may also occur.
The  phase speed of the waves can be
estimated  to be a typical sound speed in the Keplerian domain.

\section{Simulations with  initial vertical fields with net flux}\label{S5}
\noindent
The simulations described above were focussed
on conditions starting from small scale 
fields at small amplitude which lead to a well defined
turbulent state in the mean that was, within that framework, initial
condition independent. Such a situation would apply to
a non magnetic star interacting with a disk threaded by
no external flux.
However,
there may be situations where the disk
is threaded  with a large scale vertical flux. In that case the disk is not 
isolated because conducting material external
to the disk is implied as a source. The source could be the material
supplied at large  distances to the disk itself.
Concentration of the flux towards the center of the disk
may lead to a large scale poloidal field threading the
disk that could be a source of a  global outflow as has been
considered by many authors
(e.g. Blandford \& Payne, 1982, Lubow, Papaloizou \& Pringle, 1994,
Shu et al.\, 1994, Spruit, Stehle, \& Papaloizou, 1995, Konigl \& Wardle, 1996). 
Simulations of disks with net flux 
may also  be relevant to a situation in which the disk becomes
dominated by an external magnetic field  arising from the central star.
The latter point of view was adopted by Miller \& Stone (2000)
who carried out simulations of a stratified  disk in a shearing box.
They found magnetically dominated solutions.
In their case, vertical gravity was included and outflow boundary conditions
were applied in the vertical direction, but
Keplerian rotation was enforced by the boundary conditions. 
In our model this is not the case but there is no vertical stratification.
However, the  periodic boundary conditions in the vertical direction,
neither take into account any constraints arising from the external
material in which field lines might be embedded nor correctly match to an
external vacuum field. We  comment
that despite the difference in the models, the outcome
is very similar: magnetically dominated solutions.

\begin{figure}
\centerline{
\epsfig{file= 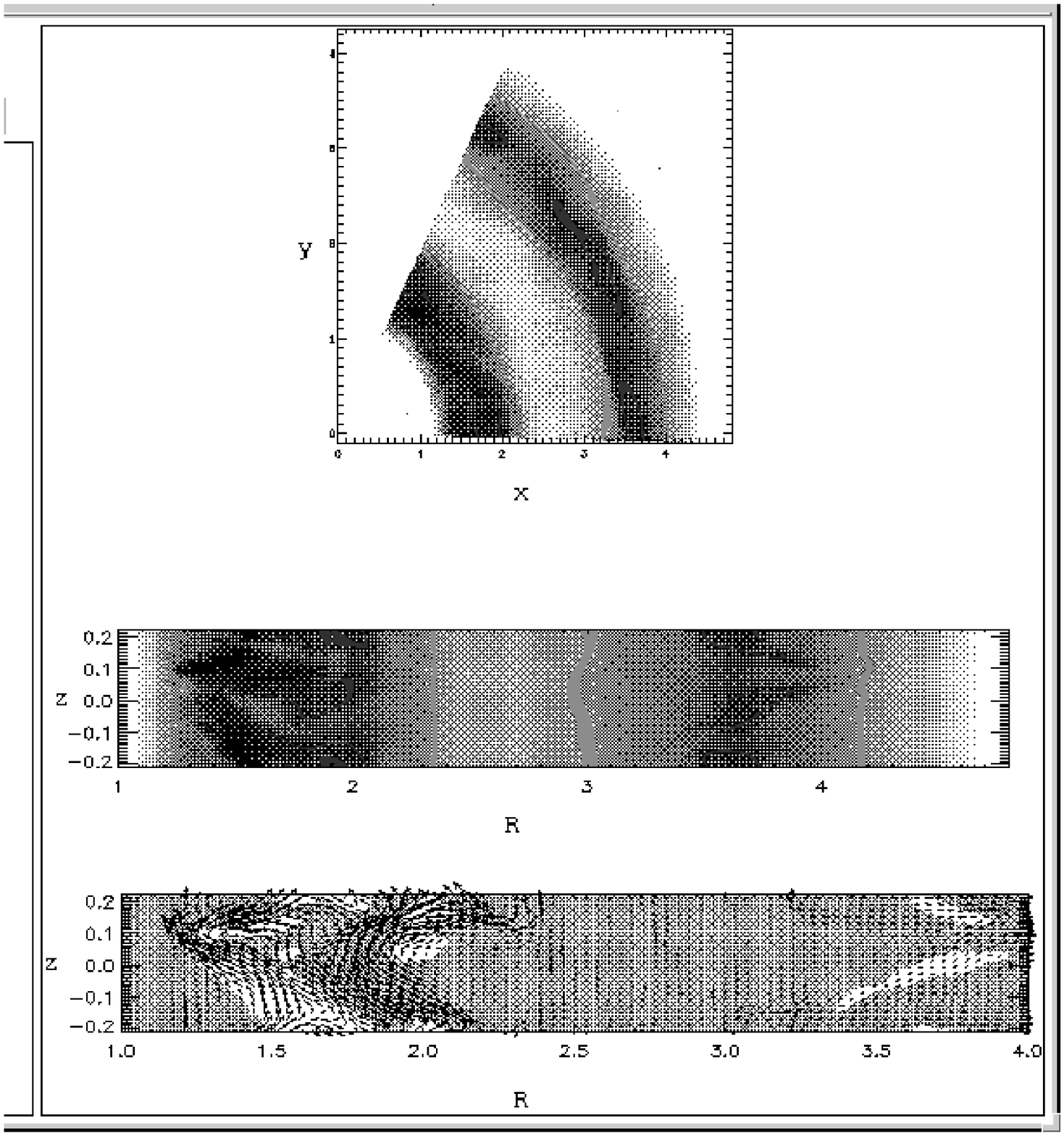,height=17.cm,width=17.cm,angle =360}}
\caption[]
{Polar contours of the surface density (upper panel) along with
the density in the  R-z-plane at $\phi=0$ and time 1224 for model n1.
The bottom panel shows the poloidal magnetic field vectors with
the superimposed binary map of the critical wavelength of the MRI.
White regions correspond to $\lambda_c$ exceeding the disk height.}
\label{fig14}
\end{figure}

\noindent
We here describe simulation n1  with initial
mean $\beta =120$ (see table \ref{table1}).
The magnetic field was applied between $R_b=1.32,$ $R_{b1}=3.72$ 
and $n_b =2.$
During the channel phase, the global magnetic energy
increases by up to 2 orders of magnitude with respect to the initial
value, and drops to a mean value of about one order of magnitude lower
than this maximum.
This model and other similar ones we ran are characterized 
by the tendency to 
produce regions with pronounced density minima or gaps
in which the  vertically 
averaged density  is smaller by 1-2 orders of magnitude
than the surroundings.
Conditions in the gaps were strongly variable, with 
inhomogeneous  variations
in $z$  and $\phi$ of sometimes more than one order of magnitude. 
This inhomogeneity is evident in figure \ref{fig14} where we  show
polar contours of the surface density (upper panel) along with a slice of the density
in the R-z-plane ($\phi$=0) at time 1224.  
 The gaps, one located next to the boundary layer,
and a second near  to the outer stable region, are visible as lumpy structures
in azimuth, and radially elongated low-density filaments alternating with
large density regions are visible in the R-z plane. This picture remains
qualitatively unchanged throughout the simulation (1444 time units).

\noindent
The formation and survival of the gaps appears to be a characteristic of
initial vertical fields with large net flux, which in turn have strong
initial channel solutions. After an associated
 reconnection, a
net vertical  magnetic flux is trapped
in low density
 regions with a radial extent larger than one local scale height.
 A large scale vertical
field 
persists in association with the gaps. In the bottom panel of figure \ref{fig14} we
show the magnetic field vectors in the R-z plane ($\phi=0$) at time 1224.
Superimposed is a binary map of the critical wavelength
{$\lambda_c$}
 (white
represents stable regions). Note that the inner gap extends roughly from 1.2 to 2,
while the outer gap is located from 3.2 to 4.
When the critical wavelength exceeds the disk height, the angular momentum transport from the faster
to the slower rotating regions
is mediated
by the torque exerted by the toroidal field that is built
up from the poloidal field.
The toroidal field is wound up until 
 torque  balance occurs maintaining the gap.
This
mechanism operates  analagous to  magnetospheric cavity formation
(e.g. Ghosh \& Lamb 1978).
In that case, a poloidal field is wound up between the disk and the central
star rotating at a different rate until
a steady torque is transmitted between the two via a force-free field.
At the  edges of the gap, the axisymmetric MRI can always operate, as the condition
for instability is fulfilled. Indeed,
figure \ref{fig14}  shows that the field stretches
out radially across the gaps until it reconnects. Filaments or blobs of magnetic
flux are expelled from both sides of the gaps, and a concentration
of vertical flux is left  there.
In this manner, the large-scale field
is maintained in the gap by the MRI.

\begin{figure}[f]
\centerline{
\epsfig{file=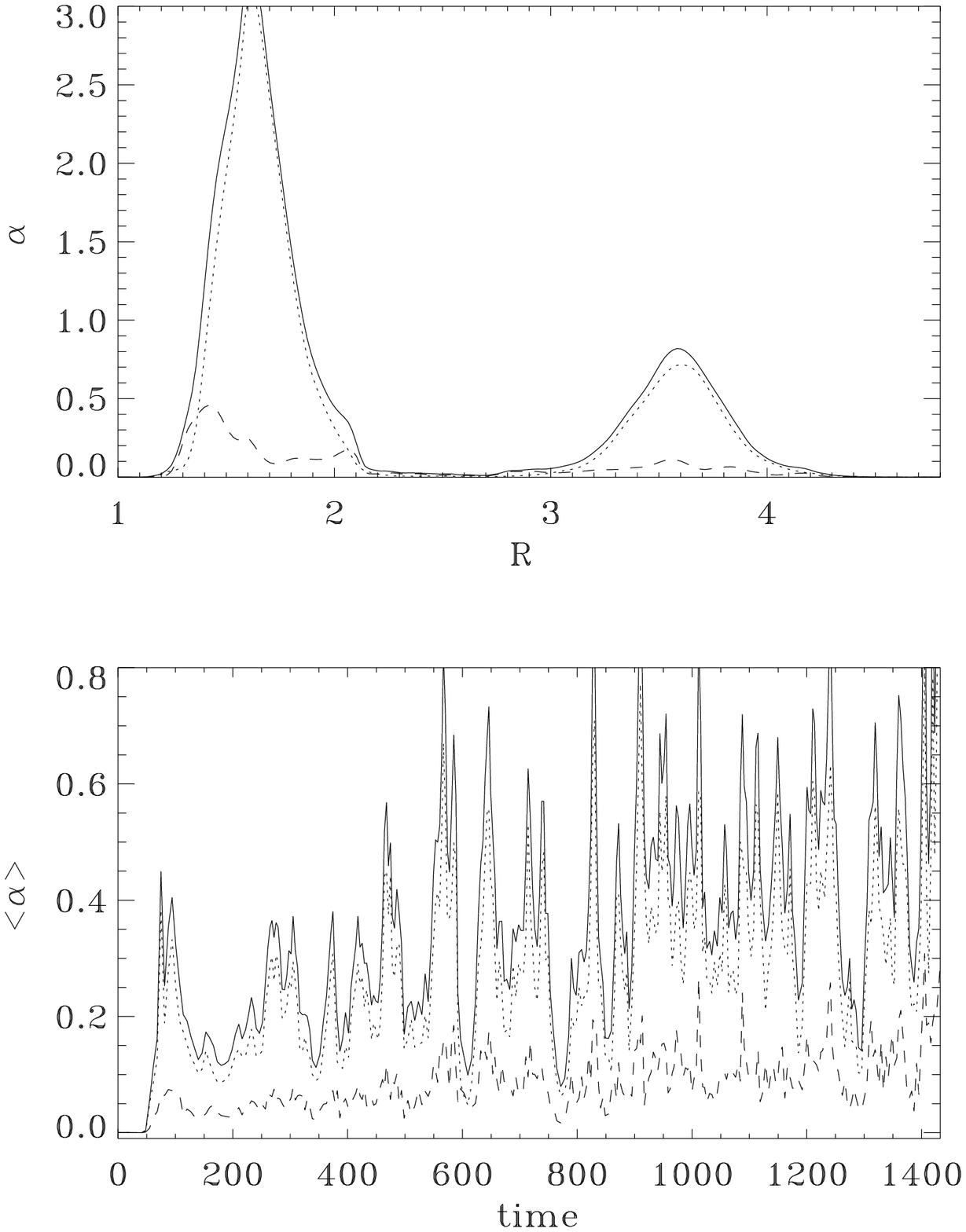,height=12.cm,width=10.cm,angle =360}}
\caption[]
{Radial
dependence of $\alpha$ for time 1143 (upper panel)
and time evolution of the volume averaged $\alpha$ (lower panel) for
run n1. The dotted curve represents the magnetic stress contribution
and the dashed curve the Reynolds stress contribution.}
\label{fig15}
\end{figure}

\noindent
We find large values of $\alpha$ correlated with the low density within
the gaps.
The upper panel of figure \ref{fig15}
shows the radial dependence of
$\alpha$ at time 1143,
while the lower panel represents the
time evolution of the volume averaged $\alpha$.
$\alpha$ periodically becomes larger than unity in the regions corresponding to
the gaps. In some models, values of $\alpha$ as large as 3 can be attained
(this finding being in agreement with Hawley (2001)).
The Maxwell stress dominates over the Reynolds stress
there, while outside the gaps, the Reynolds stress becomes important.
Outside the gaps, $\alpha$ may decrease to a mean of 0.005, similar to
the zero net field models. 

\begin{figure}[b]
\centerline{
\epsfig{file=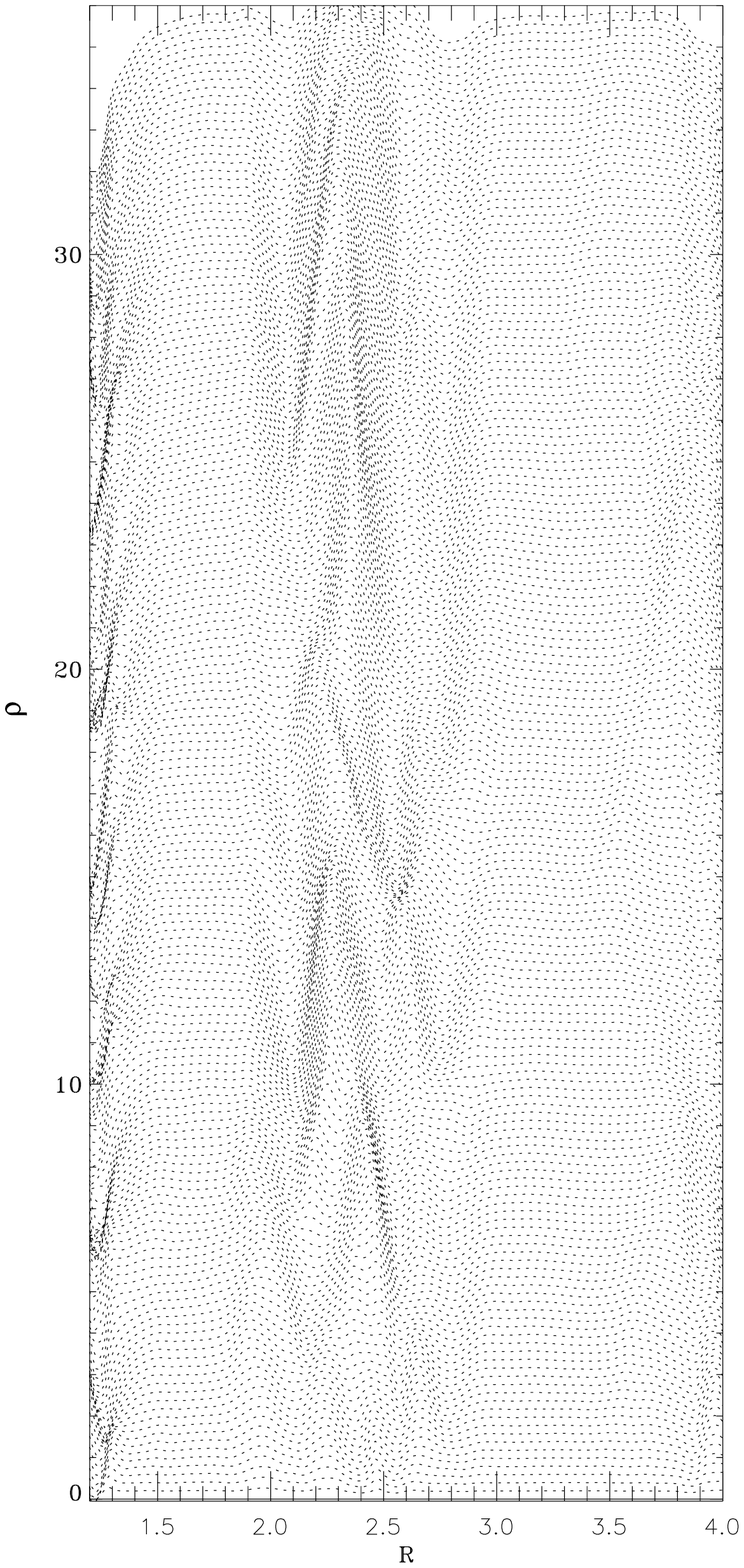,height=14.cm,width=8.cm,angle =360}}
\caption[]
{Time sequence of the averaged density for model n1 between
time 565 and 609.}
\label{fig16}
\end{figure}

\noindent
When gaps are formed
next to the boundary layer, the large-scale poloidal magnetic field trapped
in that region contributes 
to link the boundary layer to the outer disk.
A periodic broadening and narrowing of the boundary layer becomes
evident in this case. We attribute this oscillation to the interaction
between the disk and the boundary layer via the field connecting these
regions. A time sequence of azimuthally and vertically averaged density
profiles (represented in figure \ref{fig16}) shows that waves, most likely
excited by this oscillation of the boundary layer propagate
through the high density region exterior to
the gap.

\noindent
We performed a number of simulations with varying resolution (e.g. run n2) and
computational domain (e.g. n3). All these runs show the formation
of at least one prominent gap (located next to the boundary layer) and
the presence of one or two smaller gaps which have the tendency to deepen
with time and sometimes merge. Their position may vary slightly with resolution
and extent of the radial or azimuthal domain. These findings are in agreement with Hawley (2001).

\subsection{A simulation with  a toroidal field with net flux}
\begin{figure}[b]
\centerline{
\epsfig{file=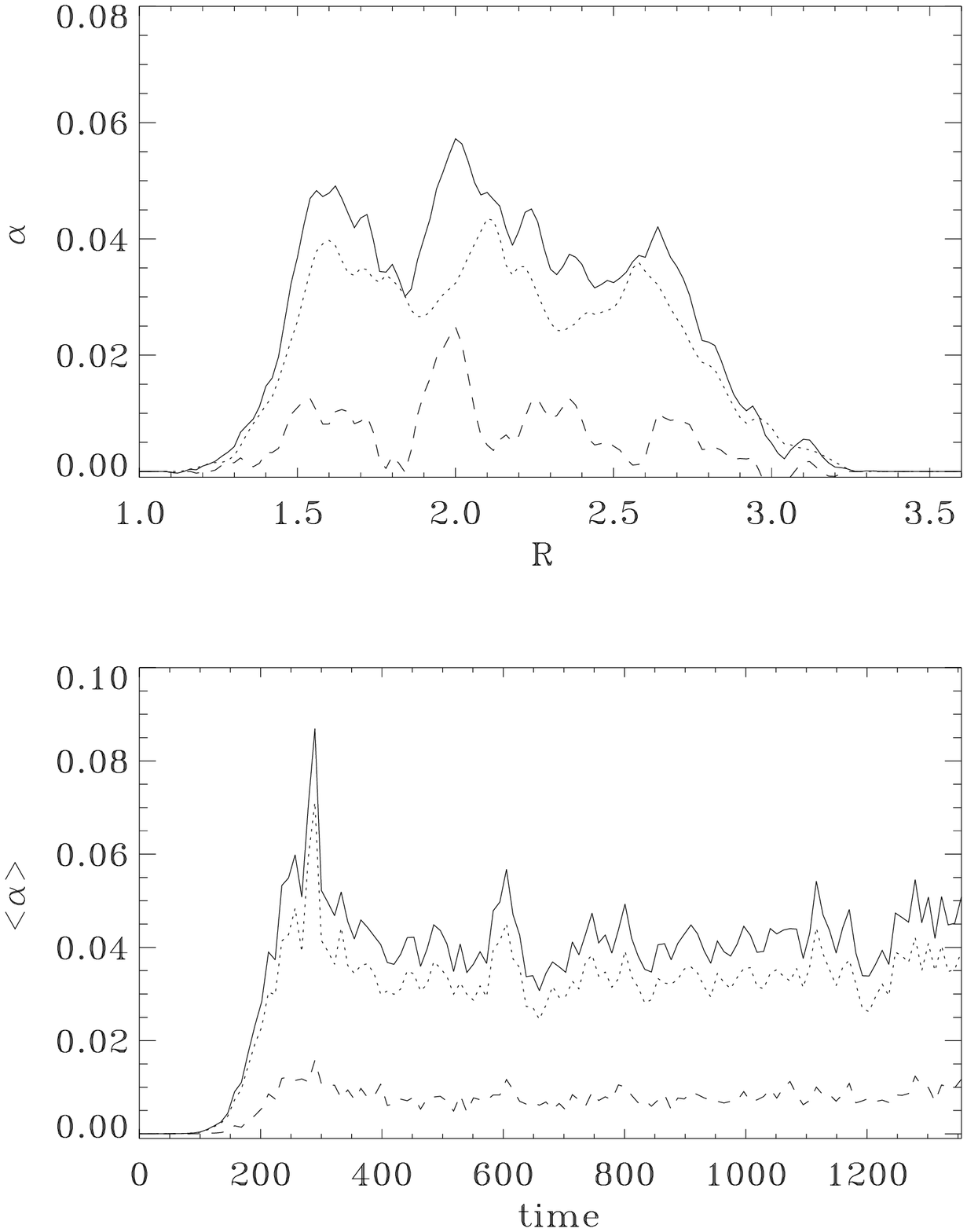,height=12.cm,width=10.cm,angle =360}}
\caption[]
{Radial
dependence of $\alpha$ for t=1388 (upper panel)
and time evolution of the volume averaged $\alpha$ (lower panel) for
run n4. The dotted curve represents the magnetic stress contribution
and the dashed curve the Reynolds stress contribution.}
\label{fig17}
\end{figure}

\noindent
Simulation  n4 was performed
with an initial toroidal field with net flux and  with one maximum in R.
The magnetic field was applied between $R_b=1.32$ and $R_{b1}=2.76$
and $n_b =2.$
As in the  cases with zero net flux,
the instability  is weak when compared to models with an initial
vertical
field. The magnetic energy first shows growth after time 125.6 
peaking  after 289 then  having grown by a factor of two. 
A quasi-steady turbulent state was reached after about 377 time units,
with the global magnetic energy
being 0.8 of the initial value at the end of the simulation (time=1444). 
The radial
variations of the vertically and azimuthally
averaged  stress parameter $\alpha$ are relatively
small, not exceeding a factor
of 3.
The lower panel of figure \ref{fig17} shows the time variation of $\alpha$ at R=2, and indicates
a mean value of 0.04. In the upper panel, we represent the radial dependence of $\alpha$
close to the end of the simulation (t=1388). The magnetic stress is always dominant,
and the Reynolds stress can sometimes become negative as previously discussed  and 
the boundary layer moves slightly inwards in the course of the simulation.

\section{Discussion}\label{S6} 
\noindent
In this paper we have studied the  time dependent evolution of a 
near Keplerian accretion
disk which 
rotates between
two bounding regions with  initial rotation profiles that are stable
to the MRI. The inner region models the boundary layer between
the disk and a central star.
 Because we  self-consistently incorporate
MRI-induced turbulence as the source of viscosity,
 the  necessity of an ad-hoc viscosity prescription 
of the type used in earlier studies  is removed.

\noindent
 For this first study we assumed the disk to be unstratified
with aspect ratio
$ H/R \simeq 0.1-0.2 $  by adopting a  cylindrically
symmetric  potential  assumed to be exclusively
generated by the central object.
 For most of the cases considered, the dynamical evolution of the disk 
following from the imposition of
different magnetic field configurations was followed over a time
span  between one hundred and two hundred rotation periods at the inner edge of
the computational domain. However, simulations of
radially more extended disks were followed up to one thousand orbital periods
at the inner edge of
the computational domain. For initial conditions
both toroidal and poloidal
magnetic fields with zero as well as net flux were  
applied in varying domains contained within the  near Keplerian
disk.

\noindent
Simulations starting from toroidal and poloidal fields with zero net flux
and a small scale of radial variation 
evolved to a  state characterized by a smooth angular velocity and
density profile similar to the 
initial one.
This was independent of the type, and within numerically
determined limits,  the amplitude of the initial field.
This result also holds in shearing box simulations (HGB96).
 But one must bear in mind
that in numerical computations, the range of initial plasma
betas to which one has reasonable access is restricted by the
resolution.
Typical values 
of $\langle \alpha \rangle$  representing a volume
average over the Keplerian  domain are 
0.004$\pm$0.002. 
Moreover, runs with a radially 
extended disk
showed  that the saturated turbulent state is maintained over more
than fifty orbital periods at the outer boundary of the Keplerian
domain.

\noindent
While the shearing box approach guarantees flux conservation for 
all time in a Keplerian domain, in global simulations this is not necessarily the case.
This is because  
in large scale simulations the
inevitable diffusion of the magnetic field out of the Keplerian
domain and into the boundary domains can lead to the 
violation  of flux conservation in the Keplerian domain.
This happens even though the boundary domains are stable to
the MRI.
  However, we found that
different prescriptions of the inner  boundary layer region  do not affect
the final state in the Keplerian domain even though the behavior
of the boundary layer itself may   vary significantly.
Once significant field has leaked into the boundary layer, toroidal
field is built up due to the shear and causes $\alpha$ to
attain negative values corresponding to inward angular
momentum transport and mass accretion. In such cases, the boundary layer expands
outwards but radial motions remain subsonic.

\noindent
We also find a similar approximate correlation
between the Keplerian domain
 volume averaged ($R \phi$)-component of the stress
and magnetic energy
as  found in  local simulations:  For simulations with zero net flux
we find, after averaging out short term variations,
 $\langle \alpha \rangle\sim  0.5/\langle \beta \rangle$.

\noindent
Models with an initial vertical field with zero-net
flux and large scale of radial variation exhibit local minima
in the density associated with maxima
of the angular velocity. These 
density pockets can reach a contrast of one order of magnitude 
with respect to the surroundings. The turbulent
state can nonetheless be characterized by an average $\alpha$
similar to the cases with small scale of radial variation.
It may require a  very  long time scale  for such states
to relax to those found in the initially small scale field cases.

 \noindent All models display large variations of
$\alpha$ in time and radius including oscillations on  
a rotational time scale.
Variations of  one order of magnitude are typical.

\noindent
 Models staring with fields that
have non zero net flux  lead to a higher level of turbulence. Thus,
model  n4  that started with a toroidal field attains 
an average alpha of 0.04.
Those starting with vertical fields with net flux such as n1
may display several
gaps in density, with the radially innermost gap typically located next to the boundary
layer.
The density contrast in the most prominent gaps can
reach up to 2 orders of magnitude
(in an azimuthal and vertical average) with respect to their surroundings.
In the gap-regions, $\alpha$ alternates between values $<1$ and
$>1$, sometimes exceeding 3 and dropping to an average of 0.005
(comparable with the zero net flux simulations) in non gap regions.
Values of $\alpha$ exceeding 1 indicate angular momentum transport by
magnetic torques originating from
fields that connect across
the gap region.

\noindent
Recognizing that there are issues to be resolved regarding
the correct boundary conditions
required to represent the effects of external conducting material,
these solutions might be relevant when the disk 
becomes dominated by an external magnetic field.
  Such magnetic fields
may affect a variety of processes that take place in accretion disks,
from dust coagulation to the interaction of planets with the disks
in which they have been formed.

\noindent
Clearly there is much room for future 
improvements and developments.
 Convergence needs 
to be checked at much higher resolution than currently attainable.
Studies of  more extended inner MRI-stable boundary layers
should be carried out, and vertical stratification
should be included. In this connection the simple periodic
boundary conditions used in the vertical direction leave the
vertical flux relatively unconstrained and unconnected
to any conductors external to the disk. Proper matching
of boundary conditions to external fields is also
an issue that may affect the behavior of the low density
gap regions studied here and a subject for future study.

\subsection{Acknowledgements} 

We would like to thank Richard Nelson for 
encouragement and support regarding
computational matters and him
together with  Caroline Terquem  and Greg Laughlin 
for valuable discussions.
We  acknowledge support from the UK Astrophysical
Fluids Facility and the NASA Advanced Supercomputing
Facility's Information Power Grid Project's Pool at
NASA ames Research Center. 
AS thanks the Astronomy Unit at QMW for
hospitality, 
 the European
Commission for support under contract number ERBFMRX-CT98-0195
(TMR network ``Accretion onto black holes, compact stars
and protostars'')
 and the NRC  for a research  fellowship. 


{}

\begin{thebibliography}{DUM}

\bibitem[]{}
Armitage, P. J., 1998, ApJ, 501, L189

\bibitem[]{}
Balbus, S. A.,  Hawley, J. F., 1991,
ApJ, 376, 214

\bibitem[]{}
Balbus, S. A.,  Papaloizou, J. C. B., 1999,
ApJ, 521, 650

\bibitem[]{}
Blandford, R. D.,  Payne, D. G., 1982, MNRAS, 199, 883

\bibitem[]{}
Blumenthal, G. R., Yang, L. T.,  Lin, D. N. C., 1984, ApJ, 287, 774    

\bibitem[]{}
Brandenburg, A., Nordlund, \AA., Stein, R. F.,  Torkelsson, U., 1995,
ApJ, 446, 741

\bibitem[]{}
Brandenburg, A., Nordlund, \AA., Stein, R. F.,  Torkelsson, U., 1996,
ApJ, 458, L45


\bibitem[]{}
Bryden, G., Chen, X., Lin, D. N. C., Nelson, R. P.,  Papaloizou,
J. C. B., 1999, ApJ, 514, 344

\bibitem[]{}
Fleming, T. P., Stone, J. M.,  Hawley, J. F., 2000, ApJ 530, 464

\bibitem[]{}
Ghosh, P.,  Lamb, F. K., 1978, ApJ, 223, L83

\bibitem[]{}
Hawley, J. F., Gammie, C. F.,  Balbus, S. A., 1995, ApJ, 440, 742

\bibitem[]{}
Hawley, J. F.,  Stone, J. M., 1995,
Computer Physics Communications, 89, 127                

\bibitem[]{}
Hawley, J. F., Gammie, C. F.,  Balbus, S. A., 1996, ApJ, 464, 690 (HGB96)

\bibitem[]{}
Hawley, J. F., 2000, ApJ, 528, 462

\bibitem[]{}
Hawley, J. F., 2001, ApJ, 554, 534 

\bibitem[]{}
Hawley, J. F.,  Krolik, J. H., 2001, ApJ, 548, 348
                                                                 

\bibitem[]{}
Kato, S., 1978, MNRAS, 185, 629

\bibitem[]{}
Kley, W., 1989, A\&A, 208, 98

\bibitem[]{}
Kley, W.,  Papaloizou, J. C. B., 1997, MNRAS, 285, 239

\bibitem[]{}
Kley, W., D'Angelo, G.,  Henning, 2001, Th., ApJ, 547, 457 

\bibitem[]{}
Konigl, A., Wardle, M., 1996, MNRAS, 279, L61

\bibitem[]{}
Lightman, A. P.,  Eardley, D. M., 1974, ApJ, 187, L1                         

Lubow, S. H., Papaloizou, J. C. B.,  Pringle, J. E., 1994, MNRAS, 267, 235

\bibitem[]{}
Lynden-Bell, D., Pringle, J. E., 1974, MNRAS, 168, 603

\bibitem[]{}
Miller, K. A., Stone, J. M., 2000, ApJ, 534, 398

\bibitem[]{}
Narayan R., 1992, ApJ, 394, 261
 
\bibitem[]{}
Nelson, R. P., Papaloizou, J. C. B., Masset, F., Kley, W., 2000, MNRAS, 318, 18

\bibitem[]{}
Papaloizou, J. C. B., Stanley, G. Q. G., 1986, MNRAS, 220, 593 (PS)

\bibitem[]{}
Popham R., Narayan R., 1992, ApJ, 394, 255

\bibitem[]{}
Pringle, J. E., 1977, MNRAS, 178, 195

\bibitem[]{}
Sano, T., Inutsuka, S.-I., Miyama, S. M., 1998, ApJ, 506, L57

\bibitem[]{}
Shakura, N. I., Sunyaev, R. A., 1973, A\&A, 24, 337

\bibitem[]{}
Shu, F., Najita, J., Ostriker, E., Lizano, S., 1994, ApJ, 429, 781

\bibitem[]{}
Spruit, H. C., Stehle, R., Papaloizou, J. C. B., 1995, MNRAS, 275, 1223

\bibitem[]{}
Stone, J. M., Hawley, J. F., Gammie, C. F., Balbus, S. A., 1996,
ApJ, 463, 656

\bibitem[]{}
Ziegler, U.,  R\"udiger, G., 2000, A\&A, 356, 1141

\bibitem[]{}

\end{thebibliography}
\end{document}